\documentclass[twocolumn]{aastex62}

\usepackage{natbib}
\usepackage{lineno}
\usepackage{amsmath}
\usepackage{subfigure}
\setcitestyle{sort,comma}
\usepackage{csvsimple}

\usepackage{rotating}
\usepackage{graphicx}
\usepackage{appendix}

\graphicspath{{./}{figures/}}

\shorttitle{Inhomogeneous Cloud Cover on NGTS-10\,A\rm{b}}

\begin{document}




\title{\large{Inhomogeneous Cloud Coverage and Altitude-Dependent Heat Transport on the Hot-Jupiter NGTS-10\,Ab from its Optical-to-Infrared Phase Curve}}

\correspondingauthor{Louis-Philippe Coulombe}
\email{louis-philippe.coulombe@umontreal.ca}

\author[0000-0002-2195-735X]{Louis-Philippe Coulombe}
\affiliation{Plan\'{e}tarium de Montr\'{e}al, Espace pour la Vie, 4801 av. Pierre-de Coubertin, Montr\'{e}al, Canada}
\affiliation{Trottier Institute for Research on Exoplanets, Department of Physics, Universit\'{e} de Montr\'{e}al, Montr\'{e}al, Canada}

\author[0000-0001-9521-6258]{Vivien Parmentier} 
\affiliation{Laboratoire Lagrange, Université de la Côte d'Azur, Observatoire de la Côte d'Azur, CNRS, Bd de l'Observatoire, Nice, France}

\author[0000-0002-7352-7941]{Kevin B. Stevenson} 
\affiliation{Johns Hopkins APL, 11100 Johns Hopkins Rd, Laurel, MD, USA}

\author[0000-0003-2278-6932]{Xianyu Tan} 
\affiliation{Tsung-Dao Lee Institute \& School of Physics and Astronomy, Shanghai Jiao Tong University, Shanghai, People's Republic of China}

\author[0000-0001-8291-6490]{Everett Schlawin}
\affiliation{Astrophysics \& Space Center, Schmidt Sciences, New York, NY 10011, USA}

\author[0000-0003-0156-4564]{Luis Welbanks}
\affiliation{School of Earth and Space Exploration, Arizona State University, Tempe, AZ, USA}

\author[0000-0003-4844-9838]{Jake Taylor}
\affiliation{Astrophysics, Department of Physics, University of Oxford, Parks Rd, Oxford, UK}

\author[0000-0003-3980-7808]{Yao Tang}
\affiliation{Department of Astronomy \& Astrophysics, University of California, Santa Cruz, CA, USA}

\author[0000-0002-2338-476X]{Mike Line}
\affiliation{School of Earth and Space Exploration, Arizona State University, Tempe, AZ, USA}

\author[0000-0001-9289-0570]{Hinna Shivkumar}
\affiliation{Anton Pannekoek Institute for Astronomy, University of Amsterdam, Science Park 904, Amsterdam, the Netherlands}

\author[0000-0003-4733-6532]{Jacob L. Bean}
\affiliation{Department of Astronomy \& Astrophysics, University of Chicago, Chicago, IL, USA}

\author[0000-0002-0875-8401]{Jean-Michel Désert}
\affiliation{Leibniz Institute for Astrophysics Potsdam, An der Sternwarte 16, Potsdam, Germany}
\affiliation{Anton Pannekoek Institute for Astronomy, University of Amsterdam, Science Park 904, Amsterdam, the Netherlands}
\affiliation{DESY, Platanenallee 6, D-15738 Zeuthen, Germany}

\author[0000-0002-9843-4354]{Jonathan J. Fortney}
\affiliation{Department of Astronomy \& Astrophysics, University of California, Santa Cruz, CA, USA}

\author[0000-0002-8518-9601]{Peter Gao}
\affiliation{Earth and Planets Laboratory, Carnegie Institution for Science, Washington, DC, USA}

\author[0000-0002-6893-522X]{Mark Hammond}
\affiliation{Atmospheric, Oceanic, and Planetary Physics, Department of Physics, University of Oxford, Parks Rd, Oxford, UK}

\author[0000-0002-1337-9051]{Eliza M.-R. Kempton}
\affiliation{Department of Astronomy \& Astrophysics, University of Chicago, Chicago, IL 60637, USA} 

\author[0000-0002-9258-5311]{Thaddeus D. Komacek}
\affiliation{Atmospheric, Oceanic, and Planetary Physics, Department of Physics, University of Oxford, Parks Rd, Oxford, UK}

\author[0000-0003-4241-7413]{Megan Weiner Mansfield}
\affiliation{Department of Astronomy, University of Maryland, College Park, MD, USA}





\begin{abstract}

Hot-Jupiters, gas giant planets with equilibrium temperatures above 1,000\,K, host large temperature gradients between their permanent day- and nightsides, resulting in circulation regimes that have no Solar System analogs. Past Kepler and Spitzer measurements have shown that the phase curves of these objects typically peak westward of the substellar point at optical wavelengths and eastward in the infrared, indicative of a possible interaction between circulation and cloud coverage. However, few hot-Jupiters have joint reflected light and thermal emission measurements, preventing a definitive statement as to the link between these phenomena. Here, we present the first phase curve of a hot-Jupiter that separates the contributions of thermal emission and reflected light through JWST observations of NGTS-10\,Ab with the NIRSpec PRISM instrument ($\lambda = 0.5$--$5.5\,\mu$m).
Using the spectrally-resolved phase curve, we jointly retrieve the planet's thermal and reflected light maps. We show that the temperature and reflectance distributions are anti-correlated, which is best explained by clouds evaporating from the eastern substellar region ($\varphi \approx -17$ to $48^\circ$) where temperatures are highest.
This is further confirmed by comparisons of the phase-resolved spectra with three-dimensional circulation models, which show that NGTS-10\,Ab's atmosphere hosts inhomogeneous cloud coverage, likely made-up of $\mu$m-sized silicate particles, and weak atmospheric drag ($\tau_\mathrm{drag}>10^6$\,s). Finally, by measuring the spectral variation of the thermal phase curve offsets,
we infer a slope of $7.1\pm1.9$\,degrees per pressure dex, indicative of heat transport that becomes more efficient at depth.
Future optical and infrared hot-Jupiter phase curve measurements over a wide range of equilibrium temperatures will enable a complete mapping of the interplay between heat transport and cloud formation in highly-irradiated exoplanet atmospheres.

\vspace{10mm}

\end{abstract}



    

 \section{Introduction} \label{sec:intro} 

Hot-Jupiters, gas giant exoplanets with orbital periods of less than 10\,days and equilibrium temperatures above 1,000\,K, host extended atmospheres amenable to precise characterization via thermal emission, reflected light, and transmission spectroscopy measurements \citep[e.g.,][]{Seager1998,Seager2000,Charbonneau2002,Charbonneau2005_thermal,Deming2005_infrared,Rowe2008}. The proximity of these planets to their host star is expected to result in a tidally-locked configuration: one hemisphere, the dayside, perpetually faces its host star and the other, the nightside, faces towards outer space and receives no irradiation \citep{Guillot1996}.

Theoretical models of hot-Jupiter climates have yielded a multitude of predictions that can be tested through observations. The irradiation gradient between the dayside and nightside leads to longitudinal and latitudinal temperature differences of several hundred Kelvins, resulting in significant variations in chemistry, cloud condensation, and vertical thermal structure across the atmosphere. The dayside's mid- and upper-atmosphere absorbs the incoming stellar flux, reducing the gradient of its vertical thermal structure \citep{Fortney2008}, whereas the nightside's temperature-pressure profile resembles that of a self-luminous body \citep[e.g.,][]{Parmentier2016}. As temperatures decrease towards the nightside, the thermal structure intersects with condensation curves of major cloud constituents (e.g., sulfides, silicates, metals; \citealt{Visscher2010,Morley_2012}), and the atmosphere may transition from a CO-dominated to CH$_4$-dominated composition in chemical equilibrium \citep{Visscher_2011}. Three-dimensional general circulation models (GCMs), which solve for radiative transfer and the primitive equations, have been adapted to hot-Jupiters to study atmospheric dynamics in this regime of extreme irradiation \citep[e.g.,][]{Showman2002,Cooper_2005}. 
A ubiquitous prediction from these models is the formation of an eastward
superrotating equatorial jet, which redistributes heat from the dayside towards the nightside through kilometer-per-second winds \citep[e.g.,][]{Showman2002,Showman2009,Heng_2011,Rauscher2012,Cho2015}. 


The atmospheric dynamics of transiting exoplanets can be probed directly through their phase curves. As the planet performs a full rotation about its own axis over one orbital period, different portions of its atmosphere come in and out of view. For thermal phase curves, the offset and amplitude are sensitive to the extent of the equatorial jet and efficiency of heat redistribution to the nightside \citep{Parmentier_2018}. When performed at multiple wavelengths, these observations constrain the longitudinal variation of the atmospheric composition, cloud properties, vertical thermal structure, and altitude-dependent changes in dynamical and radiative timescales \citep[e.g.,][]{Stevenson2014,Arcangeli2021,Jacobs2025}.

Phase curve measurements of hot-Jupiters in the last decades have occurred on two fronts: at optical wavelengths with the CoRoT \citep{CoRoT2009}, Kepler \citep{Borucki2010}, TESS \citep{Ricker_2014}, and CHEOPS \citep{Benz_2020} space telescopes \citep[e.g.,][]{Snellen2009,Demory_2011,Esteves2013,Shporer2015}, and at near-infrared/infrared wavelengths with the Hubble and Spitzer space telescopes \citep[e.g.,][]{Harrington2006,Knutson2007,Stevenson2014,Kreidberg2018,Arcangeli2019}.
In the $T_\mathrm{eq}=1400$--$1800\,$K range, Kepler revealed phase curves that are significantly shifted westwards, possibly indicative of reflective clouds confined to the western portion of the dayside \citep[e.g.,][]{Demory2013,Shporer2015,Hu_2015,Morris_2024}. Contrastingly, Spitzer phase curve observations found broad evidence for eastwards offsets,
attributed to the equatorial jet which shifts the hottest region of the atmosphere away from the substellar point \citep[e.g.,][]{Knutson_2007,Knutson2009}. These results hint at a direct interplay between circulation and cloud coverage in hot-Jupiter atmospheres. However, few planets have phase curve measurements in both reflected light and thermal emission, preventing a complete understanding of the interaction between clouds and circulation in these objects' atmospheres. Furthermore, Spitzer has found westward offsets in a subset of objects \citep[e.g., CoRoT-2\,b;][]{Dang2018}, complicating population-level comparisons with the Kepler sample, as the westward offsets observed by Kepler may not necessarily correspond to eastward offsets at longer wavelengths. More recently, TESS and CHEOPS have enabled the observation of broadband optical/near-infrared phase curves, although primarily of ultra-hot Jupiters whose thermal emission dominate even in optical bandpasses \cite[e.g.,][]{Shporer2019,Wong2020,Deline2022,Demangeon2024}.

With the optical-to-infrared coverage of NIRISS SOSS ($\lambda=0.6$--$2.8$\,$\mu$m; \citealt{Doyon2023,Albert_2023}) and NIRSpec PRISM ($\lambda=0.5$--$5.5$\,$\mu$m; \citealt{Jakobsen_2022,Birkmann_2022}) onboard the JWST \citep{Gardner2006}, simultaneous reflected light and thermal emission observations are now possible. The NIRISS SOSS phase curve of the ultra-hot Neptune LTT 9779\,b ($T_\mathrm{eq} = 1978$\,K; \citealt{Jenkins2020}) revealed an asymmetric cloud distribution over its dayside, proposed to result from a colder western dayside where highly-reflective silicate clouds condense \citep{Coulombe2025a}. Furthermore, the data yielded a thermal phase curve that is symmetric about the substellar point, indicative of short radiative timescales. The phase curve of the ultra-hot Jupiter WASP-121\,b ($T_\mathrm{eq} =2358$\,K; \citealt{Delrez_2016}) was also observed with NIRISS SOSS \citep{Splinter2025}. The data revealed a moderate Bond albedo ($A_\mathrm{B} = 0.277\pm0.016$) but low geometric albedo (3$\sigma$ upper-limit of $A_g < 0.175$) for the planet, with the significant thermal flux of WASP-121\,b even at short wavelengths complicating the isolation of the reflected light component. These two objects, however, are not part of the canonical population of hot-Jupiters. LTT 9779\,b and WASP-121\,b are significantly hotter than typical hot-Jupiters, and LTT 9779\,b is also much lighter ($M_\mathrm{p} = 0.09$\,$M_\mathrm{Jup}$) and more metal-rich \citep[e.g.,][]{ashtari2025heatrevealscloudsconceal} than Jupiter-sized objects. Optical-to-infrared observations of hot-Jupiters in the $T_\mathrm{eq} = 1000$--$2000$\,K range are needed to achieve a global understanding of their climates and to anchor the population-level trends from past Kepler and Spitzer observations.

NGTS-10\,Ab is a 1.24\,$R_\mathrm{Jup}$, 2.16\,$M_\mathrm{Jup}$ gas giant orbiting NGTS-10\,A, a K5V host star with a wide-separation ($\sim$107\,au) M dwarf binary companion \citep{McCormac_2020,Parmentier2026}. With an ultra-short orbital period of 0.77\,days, NGTS-10\,Ab has an equilibrium temperature of 1495\,K, making it a near WASP-43\,b-twin ($P = 0.81$\,days, $T_\mathrm{eq} = 1370$\,K; \citealt{Hellier_2011}) in terms of orbital and physical properties. The proximity of NGTS-10\,Ab to its host star ($\sim$1.5 Roche radii) could result in strong planet--star tidal interactions and lead to an inspiral timescale of a few tens of Myr \citep{McCormac_2020}, making this system an important target for orbital decay measurements. JWST/NIRSpec PRISM Observations of NGTS-10\,Ab revealed a notable lack of methane on its nightside \citep{Parmentier2026}, a species expected to be present in detectable quantities in chemical equilibrium. This methane non-detection, along with precisely constrained abundances of water and carbon monoxide on the day- and nightsides, provided the first definitive evidence for longitudinal chemical quenching in an exoplanet atmosphere. Retrievals on the dayside spectrum also yielded a solar atmospheric metallicity and slightly sub-solar carbon-to-oxygen ratio for NGTS-10\,Ab. Moreover, the measured phase-resolved effective temperature of NGTS-10\,Ab indicates poor heat redistribution to the nightside ($\varepsilon\approx0.34$; 0 = no redistribution and 1 = full redistribution), as well as a temperature gradient between the eastern ($\varphi = 90^\circ$) and western ($\varphi= -90^\circ$) hemispheres
of $T_\mathrm{East}$-$T_\mathrm{West}\approx$ 150\,K, suggesting eastward advection partially dampened by either nightside cloud condensation or atmospheric drag. 

In this work, we perform a complementary analysis of the observations presented in \citet{Parmentier2026}, focusing on the atmospheric dynamics, cloud coverage, and altitude-dependent heat circulation of NGTS-10\,Ab through phase curve mapping and an in-depth comparison with GCMs. From the joint optical and infrared phase curve coverage, we find evidence for the presence of an inhomogeneous cloud cover on the dayside of the planet. The data also reveal phase curve offsets that vary with pressure, with the efficiency of heat transport monotonically decreasing with altitude over the photosphere. In Section \ref{sec:method}, we describe the updated \texttt{Eureka!} reduction used in this work, as well as our light-curve fitting and phase-curve mapping methodology. We present the suite of GCMs and post-processed outputs compared to the observations in Section \ref{sec:GCMs}. Our results are presented in Section \ref{sec:results} and their implications are discussed in Section \ref{sec:discussion}. Finally, we summarize our work in Section \ref{sec:conc}.



\section{Methods}\label{sec:method}

\subsection{Data Reduction and Spectral Extraction}\label{sec:data_red}

As part of JWST GO program 2158 (P.I.: Parmentier), we observed the hot-Jupiter NGTS-10\,Ab for 20.8\,hours with NIRSpec PRISM as it performed more than a complete revolution around its host star NGTS-10\,A. The data consist of 46,539 integrations taken over three exposures, with an integration time of 1.6\,seconds, beginning 1\,h before the first secondary eclipse and ending 0.3\,hours after the second.
The relatively low brightness of NGTS-10\,A ($J_\mathrm{mag} = 12.4$; \citealt{McCormac_2020}) enables the system to be observed over the full 0.5--5.5\,$\mu$m wavelength range without saturating the PRISM detector. We re-reduce the observations presented by \citet{Parmentier2026} using an updated version of the \texttt{Eureka!} pipeline \citep[v1.2.2; ][]{Bell2022}. Specific updates include using more recent JWST calibration files (pmap: 1364) and running the jump detection step with a $4\sigma$ threshold.  We still use a 3-pixel full-width aperture size to minimize contamination from NGTS-10\,B.

After extraction of the spectroscopic light curves, we cut the first 30\,minutes (1123 integrations) at the beginning of each of the three exposures due to the presence of exponential ramps. Similar systematics after exposure breaks have been observed in other NIRSpec PRISM datasets \citep[e.g.,][]{cassese2025jwsttransitjupiteranalog,kipping2025jwsttransitjupiteranalog}. Between exposures, the detector continuously resets after every frame without accumulating any charge, potentially leading to changes in pixel sensitivity from persistence effects that gradually settle over $\sim$30-minute timescales. We produce our white light phase curve by binning all wavelengths of the NIRSpec PRISM detector from $\lambda=2$--$5.5$\,$\mu$m. Wavelengths below 2\,$\mu$m are not considered in the white light curve for two reasons: 1) the short-wavelength data could contain non-negligible reflected light signal, requiring more complex phase-curve modeling \citep[e.g.,][]{Coulombe2025a}, and 2) the light curves show stronger signs of correlated noise below 2\,$\mu$m (Figure \ref{fig:2D_timeseries}), which could bias our orbital parameter inferences.

For the spectroscopic light curves, we consider two binning schemes: 46 light curves extracted from 10-column-wide bins to produce the phase-resolved spectra, and 15 light curves extracted using set bins\footnote{With bin edges [0.5, 1.0, 1.5, 2.0, 2.3, 2.6, 2.85, 3.1, 3.3, 3.5, 3.75, 4.0, 4.25, 4.5, 5.0, 5.5]\,$\mu$m} to investigate variations in phase curve morphology at higher signal-to-noise ratio (SNR). Finally, we also produce broadband light curves at optical/near-infrared ($\lambda = 0.5$--$1$\,$\mu$m) and infrared ($\lambda=3$--$5.5$\,$\mu$m) wavelengths to perform reflected light and thermal emission phase-curve mapping, respectively. For the thermal phase curve, we consider the 3--5.5\,$\mu$m range as it is similar to the wavelengths covered by NIRSpec G395H and the 3.6/4.5\,$\mu$m Spitzer photometric filters, facilitating comparison with past observations. For all light curves, we compute their running median with a window size of 100 integrations, and replace any $>4\sigma$ outlier integration by the local value of the running median.

\subsection{Light-Curve Fitting}

We fit the broadband and spectroscopic light curves by modeling the observed time-dependent flux ($f$) as the product of the astrophysical signal ($A$) and systematics model ($S$):

\begin{equation}
    f(t) = A(t)S(t).
\end{equation}

\medskip 
\noindent 
The functional form of these two components is described in detail in the sections below.

\subsubsection{Astrophysical Model}

As the observations capture the full phase curve of NGTS-10\,Ab, the astrophysical signal ($A$) corresponds to the sum of the planetary ($F_\mathrm{p}$) and stellar flux ($F_\star$), normalized by the stellar flux:

\begin{equation}
    A(t) = \mathcal{T}(t,\Omega)\frac{F_\star(t)}{\bar{F_\star}} + \mathcal{E}(t,\Omega)\frac{F_\mathrm{p}(t)}{F_\star(t)}.
\end{equation}

\medskip 
\noindent 
The transit function $\mathcal{T}$ describes the decrease in light from the star when occulted by the planet, and is equal to unity outside of transit. As for the secondary eclipse function $\mathcal{E}$, it corresponds to the fraction of the planetary disk that is visible at a given time $t$, and is null during full secondary eclipse. The shape of these two functions depends on the orbital and physical parameters of the system $\Omega$, and are modeled using the \texttt{batman} Python package \citep{Kreidberg_2015}. We consider a circular orbit for the planet ($e = 0$), as justified by its short orbital period which corresponds to a circularization timescale of $\sim$60,000 years (eq. 2 of \citealt{Jackson_2009}; assuming $Q'_\star = 10^5$), which is considerably shorter than the age of the system. It is also unlikely that NGTS-10 B, given its 107\,au separation, is inducing an eccentric orbit for NGTS-10\,Ab.

The main sources of time-variability in the stellar flux are stellar activity (e.g., rotational variation), as well as ellipsoidal variation and Doppler boosting. Given the relatively long rotation period of NGTS-10\,A \citep[17.3\,days;][]{McCormac_2020} compared to the duration of our observations, we assume that any rotational variation of the stellar flux can be approximated as a linear slope, which is handled by the systematics model (Section \ref{sec:systematics}). As for the stellar ellipsoidal variation and Doppler boosting effects, which result from the gravitational interaction between the planet and the star \citep[e.g.,][]{Shporer_2017}, we estimate their amplitudes to be of $A_\mathrm{EV} \approx36 $\,ppm and $A_\mathrm{DB} \approx9$\,ppm at $\lambda$=0.7\,$\mu$m (via equations 4 and 7 of \citealt{Shporer_2017}). These effects are not considered in the light-curve fitting as they are significantly smaller than the planetary signal, and we therefore assume the stellar flux to be constant throughout the visit ($F_\star(t)=\bar{F_\star}$).

The planetary flux is, however, expected to vary over the course of the planet's orbit. To model this modulation, we consider a second-order sinusoid function \citep[e.g.,][]{Cowan2008,Stevenson2014}

\begin{equation}\label{eq:Fp}
    F_\mathrm{p}(t) = \sum_{n=0}^{2} F_n\cos(n[\alpha(t)-\delta_n]),
\end{equation}

\medskip
\noindent 
where $F_n$ and $\delta_n$ are the amplitude and offset of an order $n$ sinusoid. The orbital phase $\alpha$ depends on the orbital period $P$ and time $t$ relative to mid-eclipse $T_\mathrm{sec}$, such that $\alpha=2\pi(t-T_\mathrm{sec})/P$. While this functional form is a valid approximation for thermal phase curves, it does not accurately reproduce the signal that would be expected for a reflected light phase curve \citep[e.g.,][]{Cowan2013}. A model properly considering reflected light geometry is presented in Section \ref{sec:slices}. Tidal distortion of NGTS-10\,Ab could also impact the shape of its phase curve. However, we estimate the variation in projected area over the course of its orbit to be $\sim$2$\%$ (via equations 1 and 2 of \citealt{akinsanmi2023effecttidaldeformationplanetary}). This is more than an order of magnitude smaller than the variation in the planet's flux, and is thus not included in the astrophysical model.

\subsubsection{Systematics Model}\label{sec:systematics}

We employ the same systematics model as considered by \citet{Parmentier2026}, which consists of a linear trend with normalization factor $c$ and slope $v$, along with two jump parameters ($j_1,~j_2$) that allow for constant flux offsets between the three data exposures:


\begin{equation}
    S(t) = c + v(t-t_0) + j_1\Theta(t-t_\mathrm{exp,2}) + j_2\Theta(t-t_\mathrm{exp,3}).
\end{equation}

\medskip 
\noindent 
The offsets are parameterized as Heaviside step functions ($\Theta(x<0)=0,~ \Theta(x\geq0)=1$) centered at the times $t_\mathrm{exp}$ where the second and third exposures begin.

\subsubsection{White Light-Curve Fit}

\begin{figure*}
    \centering
    \includegraphics[width=1.\linewidth]{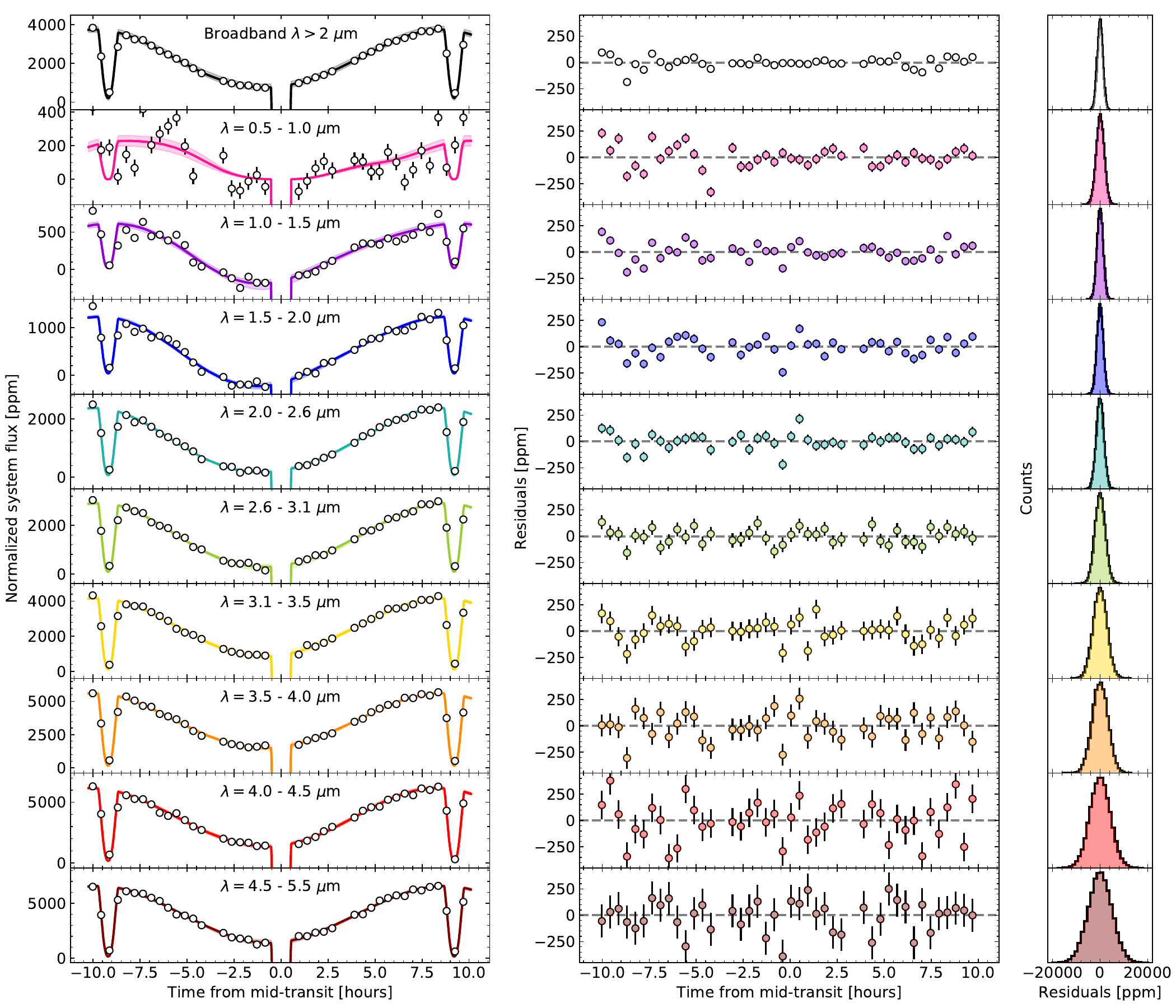}
    \caption{Broadband and spectroscopic light curve fits of NGTS-10\,Ab. Left: The systematics-corrected data, binned at intervals of 27 minutes (1,000 integrations per bin) for visual clarity, are shown as black points. The data past 2\,$\mu$m are binned into 6 light curves to display the complete wavelength coverage of NIRSpec PRISM. The median and 1-$\sigma$ confidence interval of the astrophysical model samples are shown by the lines and shaded region, respectively. The two secondary eclipses occur near the beginning and the end of the time series observations, with the primary transit occurring in-between. The modulation of the planetary flux is precisely measured in all bins, and its amplitude increases monotonically with wavelength. We show the fit performed using the $\mathcal{R}_\mathrm{cloud}$--$T_\mathrm{cond}$ Lambertian parameterization for the 0.5--1\,$\mu$m bin, and the second-order sinusoid model for the rest. Center: Residuals binned at intervals of 27 minutes, shown on the same vertical scale for all light curves. Right: Distributions of the residuals from the light curve fits, as computed from the non-binned light curves. All light curves show residual distributions that appear Gaussian.}
    \label{fig:spec_lcs}
\end{figure*}

We proceed with our white light-curve fit to constrain the system parameters that will then be fixed at the spectroscopic light-curve fitting stage. For the system parameters, we keep free the mid-transit time ($T_0$, $\mathcal{U}$[0.05, 0.15]\,BJD $-$ 2460015), the orbital period ($P$, $\mathcal{N}$[0.7668944, 0.0000003$^2$]\,days; \citealt{McCormac_2020}), the planet-to-star radius ratio ($R_\mathrm{p}/R_\star$, $\mathcal{U}$[0.1, 0.5]), the quadratic limb-darkening coefficients (LDC; [$u_1$, $u_2$], $\mathcal{U}$[-3, 3]), the scaled semi-major axis ($a/R_\star$, $\mathcal{U}$[1, 10]), and the impact parameter ($b$, $\mathcal{U}$[0, 1]). The mid-eclipse time considered in the secondary eclipse function accounts for the light time travel delay ($T_\mathrm{sec} = T_0 + P/2 + 2a/c$, where $2a/c \approx 14.7$\,s). The phase curve parameters consist of the mean planetary flux ($F_0$, $\mathcal{U}$[-10$^4$, 10$^4$]\,ppm), as well as the amplitudes ($F_n$, $\mathcal{U}$[0, 10$^4$]\,ppm) and offsets ($\delta_n$, $\mathcal{U}$[-$\pi/n$, $\pi/n$]) of the first and second harmonics ($n=1,2$). The systematics model parameters comprise the normalization factor ($c$, $\mathcal{U}$[-10$^{-9}$, 10$^9$]), the slope ($v$, $\mathcal{U}$[-10$^{-9}$, 10$^9$]\,day$^{-1}$), 
and the two jump parameters ($j_{1,2}$, $\mathcal{U}$[-5000, 5000]\,ppm). Finally, we also fit for the light curve uncertainties ($\sigma_\mathrm{phot}$, $\mathcal{U}$[1, 10$^4$]\,ppm). The light curve uncertainties are treated as a free parameter, rather than assuming the uncertainties returned by the data reduction. This is done to allow for error inflation and to marginalize our parameter constraints over any potential leftover systematics in the data and/or an inability of the astrophysical model to fully reproduce the observations. We explore the parameter space using the affine-invariant Markov chain Monte Carlo (MCMC) sampler \texttt{emcee} \citep{Foreman_Mackey_2013} with 68 walkers (4 per free parameter, for a total of 17 free parameters) and iterate for 100,000 steps. The posteriors of the fit parameters are constructed from the last 40,000 steps of MCMC, with the rest discarded as burn-in. For all MCMC analyses presented in this work, we visually inspected the chains to ensure that convergence was achieved within the burn-in phase. The white light-curve fit is shown in Figures \ref{fig:spec_lcs} and \ref{fig:wlc_fit}.

\begin{table}[hbt!]
\caption{Measured and derived orbital and physical parameters of NGTS-10\,Ab from its NIRSpec PRISM white light curve. We assume stellar parameter values of $R_\star = 0.70\pm0.03\,R_\odot$ and $T_\mathrm{eff,\star}=4567\pm36\,$K \citep{Parmentier2026} to produce the derived parameters. The equilibrium temperature is computed assuming zero Bond albedo ($A_\mathrm{B}=0$) and full heat redistribution ($f=1$). The planetary radius is derived in units of Jupiter's equatorial radius ($R_\mathrm{Jup} = 71,492$\,km).}
\centering
\begin{tabular}{lcc}
\hline
\hline
Parameter     & Measurement         \\
\hline
Transit time $T_0$ [BJD - 2460015]    &   0.082427 $\pm$ 1.6$\times10^{-5}$    \\
Period $P$ [days] & 0.76689440 $\pm$ 2.9$\times10^{-7}$ \\
Planet-to-star radius ratio $R_\mathrm{p}/R_\star$   &  $0.1815\pm0.0050$  \\
Scaled semi-major axis $a/R_\star$   & $4.53\pm0.15$ \\
Impact parameter $b$  &  $0.854\pm0.017$  \\
\hline
\hline
Derived  & \\
\hline
Planetary radius $R_\mathrm{p}$ [$R_\mathrm{Jup}$]  & $1.234\pm0.063$ \\
Semi-major axis $a$ [AU] & $0.01476\pm0.00079$ \\
Inclination $i$ [deg]  & $79.14\pm0.40$ \\
Equilibrium temperature $T_\mathrm{eq}$ [K] & $1516\pm27$ \\
\hline
\label{table:wlc_fit}
\end{tabular}
\end{table}

Our constraints on NGTS-10\,Ab's orbital parameters from its white-light curve fit are listed in Table \ref{table:wlc_fit}.
We measure a planet-to-star radius ratio of $0.1815\pm0.0050$, corresponding to a planetary radius of $R_\mathrm{p} = 1.234\pm0.063$\,$R_\mathrm{Jup}$. Our measurements of $R_\mathrm{p}/R_\star$, $a/R_\star$, and $i$ are consistent at 0.09--0.49$\sigma$ and 0.08--0.64$\sigma$ with the measurements of \citet{McCormac_2020} and \citet{Parmentier2026}, respectively. Notably, our uncertainties on these parameters are 7--14 times larger than those of \citet{Parmentier2026}, despite being derived from the same dataset. There are two reasons for this difference in precision: 1) we only consider wavelengths $>2\,\mu$m for our white light curve, resulting in a factor $\sim$1.4 increase in the photometric scatter compared to the $\lambda=0.5$--$5.5$\,$\mu$m light curve, and 2) the LDCs are kept free in our fit, whereas they were fixed to model values in \citet{Parmentier2026}. Given the high impact parameter of NGTS-10\,Ab ($b\sim0.85$), the correlation between the stellar intensity profile and orbital parameters is significant, leading to a more conservative estimate of the uncertainties when keeping the LDCs free.

From the broadband light curve fit, we infer a phase curve amplitude of $2980\pm50$\,ppm, which corresponds to an expected modulation of $\sim$125\,ppm that would be induced by tidal distortion for a Jupiter-like Love number \citep{akinsanmi2023effecttidaldeformationplanetary}. Such a modulation is smaller than the 3$\sigma$ precision that is achieved for the phase curve amplitude, indicating that this effect cannot be detected from the observations and justifying our choice to neglect it.

Another quantity of interest for NGTS-10\,Ab is its mid-transit time, as the orbital period of the planet could be steadily decreasing with time due to tidal interactions with its host star \citep{McCormac_2020}. As described in Section \ref{sec:orb_decay_analysis} of the Appendix, we perform an orbital decay analysis of NGTS-10\,Ab using archival transit time measurements \citep{Griffiths2026} along our precise (1.4\,s) JWST/NIRSpec transit time. Our analysis constrains NGTS-10\,Ab's decay rate to $\dot{P}=-4.6\pm4.2$\,ms\,yr$^{-1}$, corresponding to a non-detection of orbital decay.

\subsubsection{Spectroscopic Light-Curve Fit}

For the spectroscopic light-curve fits, we repeat the methodology of the broadband fit while keeping the orbital parameters fixed to their maximum-likelihood white-light values ($T_0$ = 2460015.082430\,BJD, $P$ = 0.76689446\,days, $a/R_\star$ = 4.71, and $b$ = 0.832). We fit for the planetary radius ($R_\mathrm{p}/R_\star$) and limb-darkening coefficients ([$u_1$, $u_2$]), the phase curve parameters ($F_n$ and $\delta_n$), and the systematics parameters ($c$, $v$, 
$j_1$, $j_2$, and $\sigma_\mathrm{phot}$), totaling 13 free parameters. The priors considered for the spectroscopic fits are identical to those of the white light curve. We perform the fits for both the 15- and 46-bin spectra, again exploring the parameter space with \texttt{emcee} using 52 walkers and iterating for 100,000 steps. For the 0.5--1\,$\mu$m bin, we consider additional phase curve parameterizations which are outlined in Section \ref{sec:refl_light_mods}. The spectroscopic light-curve fits for the low-resolution spectrum are shown in Figure \ref{fig:spec_lcs}. 

From these fits, we extract the planet-to-star flux ratio by computing the values of $F_p$ (eq. \ref{eq:Fp}) from the samples of the phase curve parameters at orbital phases -180$^\circ$ (nightside), -90$^\circ$ (evening), 0$^\circ$ (dayside), and 90$^\circ$ (morning). We also extract the phase curve offset at each wavelength. We define the offset as $\delta = -2\pi(T_\mathrm{sec}-t_\mathrm{max})/P$, where $t_\mathrm{max}$ is the time at which the phase curve reaches its peak, such that a positive value corresponds to an eastward offset. 



Upon inspection of the 46-bin phase-resolved spectra, we observe that the nightside $F_\mathrm{p}/F_\star$ spectrum is consistently $>1\sigma$ below zero for wavelengths $\lambda\lesssim1.5$\,$\mu$m (Fig. \ref{fig:FpFs_spectra_4_phases}). This dip to negative values at short wavelengths is possibly caused by a loss of continuum around the transit due to the two exposure breaks, or from leftover long-term systematics in the data. Negative nightside flux values were not observed by \citet{Parmentier2026}; this difference most likely arises from the changes in the reduction steps described in Section \ref{sec:data_red}. To address this,
we subtract the weighted median of the data points at wavelengths $\lambda<1.05\,\mu$m (-250\,ppm, in $F_\mathrm{p}/F_\star$-space) from the nightside spectrum, such that it is anchored to 0 at short wavelengths where the planetary flux is expected to be negligible.
Indeed, 
the equivalent $\lambda=0.5$--$1.05\,\mu$m weighted-median planet-to-star flux ratio of a $T_\mathrm{eff} = 1200$\,K blackbody is of $\sim$2\,ppm, thus justifying our approach. As shown in Figure \ref{fig:nightside_wo_offset}, the addition of an offset to the nightside spectrum mostly impacts the short wavelength data when shown in brightness-temperature space, as the relative change in planet-to-star flux ratio introduced by the constant $F_\mathrm{p}/F_\star$ offset decreases sharply with wavelength. We proceed with the shifted nightside spectrum for the remainder of our interpretation and further discuss the impact of this choice on the physical interpretation in Section \ref{sec:phase_spectra}. While the negative flux values at short wavelengths could impact the inferred phase curve offsets, we also perform fits in Section \ref{sec:refl_light_mods} using parameterizations that force the flux to be everywhere-positive, allowing us to assess the robustness of our measurement.



\subsection{Phase-Curve Mapping}\label{sec:slices}

We infer the longitudinal distributions of the top-of-atmosphere reflectance (hereafter simply referred to as reflectance) and brightness temperature of NGTS-10\,Ab by performing fits using functional forms specifically suited for reflected light and thermal emission phase curves.
We describe in the sections below the methodology employed for this analysis.

\subsubsection{Thermal Emission}
For the thermal emission fit, we consider a set of 18 longitudinal slices of 10$^\circ$ width. The thermal emission (in units of $F_\mathrm{p}/F_\star$) of each individual slice is a free parameter, similar to the method presented by \citet{Knutson_2007}. At a given orbital phase $\alpha$, the disk-integrated emission from the planet is equal to

\begin{equation}
    \frac{F_\mathrm{p,therm}}{F_\star} = \sum_{k=1}^{K} T_k \cdot \mathcal{K}_{\mathrm{therm},k}(\alpha),
\end{equation}

\medskip 
\noindent 
where $K=18$ is the number of slices and $T_k$ is the thermal emission value of slice $k$. The thermal emission kernel $\mathcal{K}_{\mathrm{therm},k}$ of a slice $k$ is computed by integrating its visibility over its solid angle \citep{Cowan2013,Coulombe2025a}. We perform the thermal emission fit on the 3--5.5\,$\mu$m broadband light curve, where the possible contribution from reflected light is negligible. The free parameters for this fit include the thermal emission of the 18 slices ($T_k$, $\mathcal{U}[-100,50000]$\,ppm), the planet-to-star radius ratio ($R_\mathrm{p}/R_\star$), the quadratic LDCs ([$u_1$, $u_2$]), and the systematics model parameters ($c$, $v$, $j_{1,2}$, and $\sigma_\mathrm{phot}$). Moreover, we include a regularization term $\pi_{T\mathrm{,map}}$ in the prior to avoid nonphysical map solutions with large oscillations on short spatial scales, such that


\begin{multline}\label{eq:reg_prior}
    \ln\pi_{T\mathrm{,map}} = -\frac{K}{2}\ln(2\pi\gamma_T^2) \\ - \frac{1}{2}\sum_{k=1}^{K}\left(\frac{K^2\left[T_{k+1}-2T_k + T_{k-1}\right]}{4\pi^2\gamma_T}\right)^2,
\end{multline}  

\medskip
\noindent 
where $\gamma_T$ (with units of ppm\,rad$^{-2}$) is the smoothness factor. The smoothness factor dictates the magnitude of the penalization against maps with nonzero second derivatives. Large values of $\gamma_T$ correspond to weak penalization. This is similar to the regularization terms commonly used in exoplanet and brown dwarf atmospheric retrievals, where the second derivative of the temperature-pressure profile is penalized against to ensure smooth solutions \citep[e.g.,][]{Line_2015,Pelletier_2021}. For values of $k-1$ and $k+1$ that are $<1$ and $>K$, respectively, we instead consider indices $K$ and $1$ to ensure continuity of the map at the antistellar point. We keep free the value of $\log\gamma_t$ ($\mathcal{U}[-8,8]$) in the phase curve fit, for a total of 27 free parameters, and iterate for 20,000 steps with \texttt{emcee}. 

\subsubsection{Reflected Light}\label{sec:refl_light_mods}

For the reflected light analysis, we use the 0.5--1\,$\mu$m broadband light curve where the contribution of reflected light to the planetary flux is expected to be important (e.g., $>50\%$ at $\lambda<0.8\,\mu$m for the $r_\mathrm{part}=0.5\,\mu$m cloudy GCM presented in Section \ref{sec:GCMs}). Similarly to the thermal emission phase curve, we model the disk-integrated reflected light as a function of phase via

\begin{equation}
    \frac{F_\mathrm{p,refl}}{F_\star} = \sum_{k=1}^K \mathcal{R}_k\cdot \mathcal{K}_{\mathrm{refl},k}(\alpha),
\end{equation}

\medskip 
\noindent 
where $\mathcal{R}_k$ is the reflectance of slice $k$ and $\mathcal{K}_{\mathrm{refl},k}$ is its reflected light kernel (assuming a Lambertian surface; \citealt{Cowan2013}).

We first perform a fit of the broadband reflected light phase curve using the slice method. The methodology is the same as that applied to the thermal emission phase curve, except that the value fitted for each slice is the reflectance ($\mathcal{R}$, $\mathcal{U}[0, 3/2]$; $\mathcal{R} = 3/2$ corresponds to a geometric albedo of 1). As for the prior, we employ a regularization term $\pi_{R,\mathrm{map}}$, similar to that of Equation \ref{eq:reg_prior}, which depends on the smoothness factor $\gamma_R$ with units of rad$^{-2}$. Using this parameterization, we perform three distinct fits, fixing the smoothness factor to values of $\gamma_R$ = 0.1, 1, and 10. When keeping the smoothness factor free, we find that it simply goes towards the lower boundary of the prior and produces a reflectance distribution similar to that of the $\gamma_R$ = 0.1 fit.

As we later present in Section \ref{sec:refl_therm_maps}, we find that the temperature and reflectance distributions inferred from the slice method are anti-correlated, with a hotter eastern dayside and a more reflective westward dayside. This hints at a scenario where the clouds evaporate over the eastern dayside as temperatures become sufficiently elevated. Given this, we devise a toy model to infer physically-meaningful constraints on the cloud properties of NGTS-10\,Ab. We utilize the temperature distribution inferred from the thermal emission fit and consider a model where the atmosphere is cloudy below a given condensation temperature $T_\mathrm{cond}$, such that the reflectance as a function of longitude is described as

\begin{equation}
    \mathcal{R}(\varphi) = 
     \begin{cases}
       \mathcal{R}_\mathrm{cloud}, &\text{if}~~ T(\varphi)\leq T_\mathrm{cond}\\
       0, &\text{if}~~ T(\varphi)>T_\mathrm{cond}
     \end{cases}~, 
\end{equation}

\medskip 
\noindent 
where $\mathcal{R}_\mathrm{cloud}$ is the cloud reflectance, and $T(\varphi)$ is the longitudinal temperature distribution of the planet. For temperatures above $T_\mathrm{cond}$, the atmosphere is considered cloud-free and non-reflective ($\mathcal{R}=0$). The temperature distribution $T(\varphi)$ considered for the model is the median brightness temperature map inferred from the thermal emission fit. 

The reflected light phase curve is modeled using a map with a longitudinal resolution of 4$^\circ$ (90 slices), whose reflectance distribution is dictated by the condensation temperature ($T_\mathrm{cond}$, $\mathcal{U}[0,2000]$\,K) and cloud reflectance ($\mathcal{R}_\mathrm{cloud}$, $\mathcal{U}[0,3/2]$). We also fit for the transit ($R_\mathrm{p}/R_\star$, [$u_1,u_2$]) and systematics model parameters ($c$, $v$, $j_{1,2}$, $\sigma_\mathrm{phot}$). We use 40 walkers (four per free parameter) and iterate for 20,000 steps with \texttt{emcee} to produce the posterior distributions of $T_\mathrm{cond}$ and $\mathcal{R}_\mathrm{cloud}$. We show the $\mathcal{R}_\mathrm{cloud}$--$T_\mathrm{cond}$ fit in Figure \ref{fig:spec_lcs}, and use the offset derived from this analysis for the comparison with GCMs presented in Section \ref{sec:results}. A summary of our process for the broadband, thermal, and reflected light phase curve fitting is presented in the form of a flowchart in Figure \ref{fig:flowchart} of the Appendix.


As described in detail in Section \ref{sec:refl_light_models} of the Appendix, we also perform fits of the reflected light phase curve using a shifted Lambertian model as well as the $\mathcal{R}_\mathrm{cloud}$--$T_\mathrm{cond}$ model assuming an isotropic kernel instead of the Lambertian one considered above. These fits are done in addition to the one assuming a second-order sinusoid
performed as part of the spectroscopic light-curve fitting.



\section{General Circulation Models}\label{sec:GCMs}

We run a suite of GCMs to compare to the measured phase-resolved spectra and phase curve offsets. Subsets of these models include the effects of atmospheric drag and cloud condensation with radiative feedback. We detail the GCM setup and post-processing steps in the sections below.

\subsection{The General Circulation Model: \texttt{ADAM}}


We apply the ADvanced Atmospheric MITgcm (\texttt{ADAM}) to simulate the atmospheric circulation and global cloud distributions of NGTS-10 Ab. \texttt{ADAM} is an umbrella term for exoplanet atmospheric modeling frameworks built upon the MITgcm \citep{adcroft2004} as a dynamical core and the non-grey radiative transfer of \cite{marley1999} as a main physical driver, including a lineage of models descending from the developments initiated by \cite{Showman2009} and extended through a series of key advancements and a wide range of applications afterwards \citep[e.g.,][]{lewis2010,kataria2013,parmentier2013,komacek2017,tan2019uhj,tan2021bd1,Parmentier_2020,komacek2022,steinrueck2023,lai2026,Mehta2026}. Under this naming convention, model configurations are referred to by specifying the relevant physical modules, and this work applies the active cloud tracer module to simulate clouds in a dynamically and radiatively self-consistent manner. 
The dynamical core solves the standard primitive equations of dynamical meteorology on the cubed-sphere grid, including equations of the horizontal angular momentum conservation, hydrostatic balance in the radial direction, mass continuity, thermodynamics and ideal-gas equation of state for hydrogen-helium dominated atmospheres \citep{adcroft2004}. Radiative heating and cooling by stellar irradiation and thermal fluxes of the planet are calculated through the radiative transfer model originally developed in \citet{marley1999}, revised for hot-Jupiters in \citet{Fortney2006}, and coupled to the MITgcm in \citet{Showman2009}, assuming equilibrium gas chemistry.

Clouds are simulated using tracer equations. For each condensing species, we assign a pair of tracer equations: one represents the mass mixing ratio of cloud particles, and another represents the ``vapor'' form \citep{tan2021bd1,Mehta2026}. Once condensation occurs, the vapor tracer turns into the cloud tracer, and cloud particles are assumed to be in a log-normal distribution to mimic the broad distribution driven by condensation and collision growth \citep[e.g.,][]{gao2021}:

\begin{equation}
    n(r) = \frac{\mathcal{N}_c}{\sqrt{2\pi}\sigma r}\exp \left(-\frac{[\ln (r/r_{\rm part})]^2}{2\sigma^2}\right),
\end{equation}

\medskip
\noindent 
where $\mathcal{N}_c$ is the total cloud particle number per unit mass, $n(r) = d\mathcal{N}_c/dr$ is the number density distribution, $\sigma$ is a nondimensional width and $r_{\rm part}$ is the reference particle radius. We assumed a fixed $\sigma=0.5$ to represent the potentially wide size distribution \citep{ackerman2001}. The reference radius $r_{\rm part}$ is a free parameter to control the clouds' optical depth and settling flux. Clouds decouple from the gas in the vertical direction with terminal speeds as a function of temperature, pressure, and particle size \citep{ackerman2001,parmentier2013}. The settling mass flux of clouds is computed by summing over the flux at each particle radius bin over the size distribution. Once clouds are on the hot side of the condensation curves, evaporation occurs, and cloud tracers turn to vapor tracers. Opacity by the spatial- and time-dependent clouds is included in the radiative transfer calculations. The extinction optical depth, single-scattering albedo, and scattering asymmetry factor of clouds as a function of wavelength based on the spatial and time-dependent cloud tracer are interpolated from grids of pre-existing Mie calculations of homogeneous, spherical particles \citep{ackerman2001}. 
Following models presented in our recent works \citep{Parmentier_2020,bell2024nightsidecloudsdisequilibriumchemistry,lai2026,Mehta2026}, we include three condensing species commonly expected in the temperature range relevant to NGTS-10 Ab: MgSiO$_3$, MnS, and Na$_2$S. 

In addition to clouds, our GCM implements an idealized drag in the horizontal momentum, -$\mathbf{u}/\tau_{\rm drag}$, where $\mathbf{u}$ is the horizontal velocity, to mimic dissipation of large-scale horizontal circulation by processes such as the magnetohydrodynamic effects \citep[e.g.,][]{Perna2010,rogers2014komacek}. Drag usually reduces wind speeds, increases phase-curve amplitudes, and decreases the phase-curve offsets. Here, $\tau_{\rm drag}$ is a characteristic drag timescale which is held fixed throughout the atmosphere, and we treat it as a free parameter in this work. 

We consider five distinct GCMs for our analysis: a cloudless and dragless ($\tau_\mathrm{drag} = \infty$) model (hereafter referred to as the ``fiducial'' GCM), two dragless cloudy models with reference particle sizes of $r_\mathrm{part} = 0.5$ and 5\,$\mu$m, and two cloudless models with drag timescales of $\tau_\mathrm{drag} = 10^5$ and $10^6$\,s. 

\subsection{Post-Processing}

Once the GCMs are integrated to nearly equilibrium states near the thermal photosphere, the outputs from the GCMs are post-processed by additional radiative transfer models to generate phase-dependent spectra at a higher spectral resolution for comparison with the observations, which consist of the phase-resolved planet-to-star flux ratio spectra and the spectroscopic phase curve offsets. The calculation of the thermal and reflected light components follow the method described in \citet{Fortney2006} and \citet{Parmentier2016}. We solve the two-stream radiative transfer equations along the line of sight for each atmospheric column and for each planetary phase considering absorption, emission, and scattering. The method naturally takes into account geometrical effects such as limb darkening. For the reflected light, we make the assumption of isotropy of the outgoing flux. This was verified in \citet{Parmentier2016} to lead to reasonable estimate of the albedo even in the case of asymmetric scattering, as long as the asymmetry parameter stayed below 0.9. The opacity structure is the same between the GCM and the post-processing, which is important as it ensures consistency between the thermal structure and the spectrum of the planet.

We extract the GCM spectra at phases $-180^\circ$, $-90^\circ$, $0^\circ$, and $90^\circ$ for comparison with the observed phase-resolved spectra. As for the spectroscopic phase curve offsets, they are produced by supersampling the phase-resolved GCM spectra, which have an orbital phase resolution of 10$^\circ$, at a resolution of 0.1$^\circ$ using cubic interpolation.

The dependency of the measured phase curve offsets with pressure is assessed using NGTS-10\,Ab's dayside contribution function, produced from the best-fit model of the ScCHIMERA retrieval presented in \citet{Parmentier2026}. The contribution function $\mathcal{C}$ describes the contribution of each atmospheric layer to the observed thermal emission $F(\lambda)$, and is computed via its Jacobian $\mathcal{C}(p,\lambda) = \partial F(p,\lambda)/\partial T(p)$ (where $p$ is pressure). The pressure level probed by a given spectroscopic offset measurement is then assumed to be the pressure at which the contribution function is highest for the observed wavelength.


\section{Results}\label{sec:results}

We present in the sections below the results from the phase-curve fitting and comparison of the observations with GCMs.

\subsection{Phase-Resolved Spectra}\label{sec:phase_spectra}

\begin{figure*}
  \centering  \includegraphics[width=1.0\textwidth]{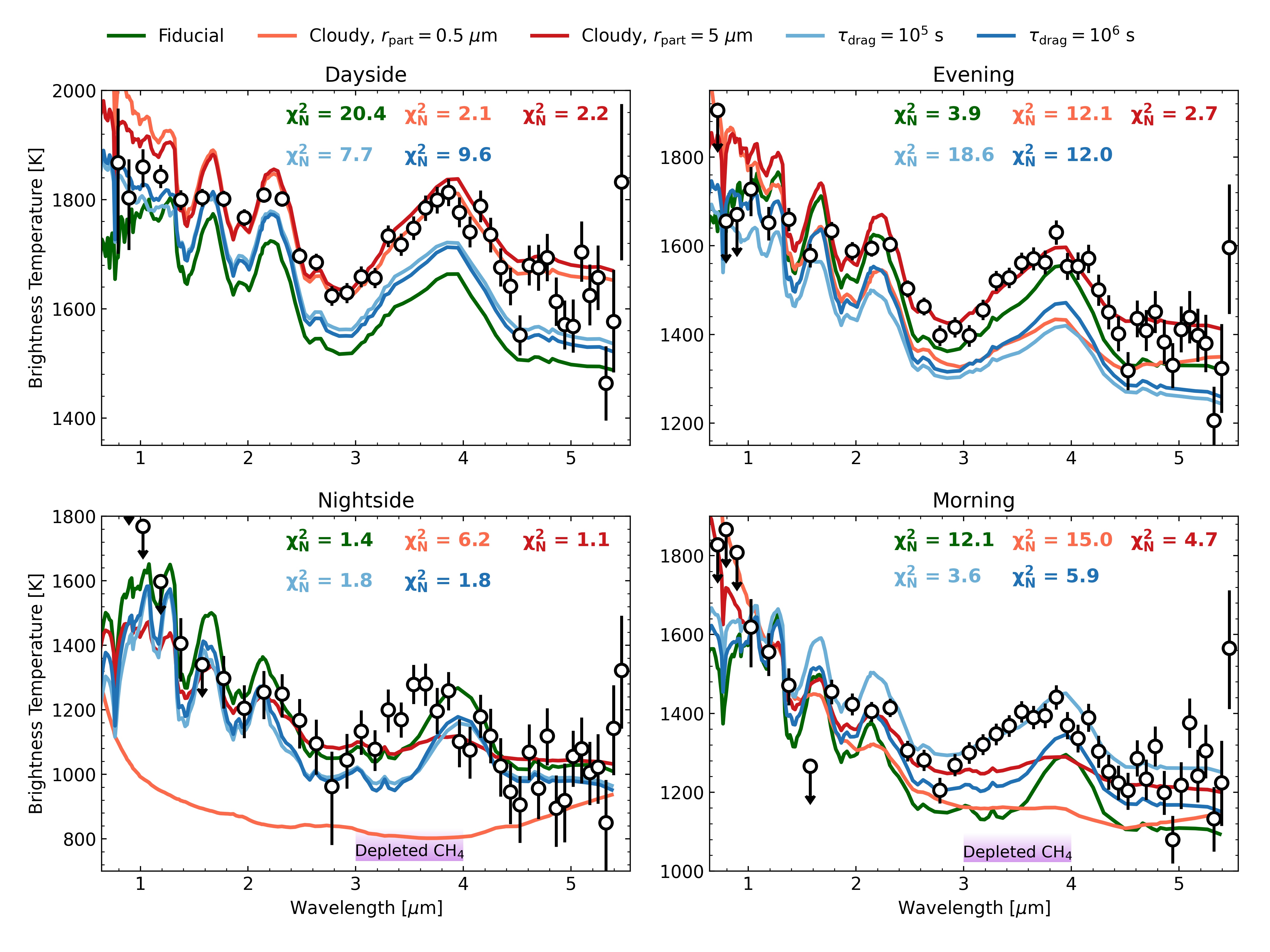}
  \vspace{-8mm}
  \caption{Brightness temperature spectra of NGTS-10\,Ab (black circles) extracted at orbital phases -180$^\circ$ (nightside), -90$^\circ$ (evening), 0$^\circ$ (dayside), and 90$^\circ$ (morning). Data that are less than 1$\sigma$ away from zero in $F_\mathrm{p}/F_\star$-space are shown as 2$\sigma$ brightness temperature upper limits. The data are compared to phase-resolved spectra from the fiducial GCM (green), cloudy $r_\mathrm{part} = 0.5\,\mu$m (orange) and $r_\mathrm{part} = 5\,\mu$m (red) GCMs, as well as the $\tau_\mathrm{drag} = 10^5$\,s (light blue) and $\tau_\mathrm{drag} = 10^6$\,s (blue) models. The reduced chi-squared values between the models and data, as computed from the $F_\mathrm{p}/F_\star$ spectra, are indicated for each orbital phase. Strong drag ($\tau_\mathrm{drag}<10^5$\,s) or nightside clouds are needed to explain the elevated dayside temperature of NGTS-10\,Ab. For the nightside spectrum, clouds with small particle sizes are disfavored, as they would result in a low-temperature, blackbody-like spectrum. The total reduced chi-square of the model fits to the four phase-resolved spectra are $\chi^2_{N\mathrm{,tot}}$ = 9.5 for the fiducial model, 8.9 for $r_\mathrm{part}$ = 0.5\,$\mu$m, 2.7 for $r_\mathrm{part}$ = 5\,$\mu$m, 7.9 for $\tau_\mathrm{drag}$ = 10$^5$\,s, and 7.3 for $\tau_\mathrm{drag}$ = 10$^6$\,s.}
  \label{fig:FpFs_4_phases}
\end{figure*}

The measured phase-resolved spectra are compared to the GCMs, which include the contributions of reflected light and thermal emission, in Figure \ref{fig:FpFs_4_phases}. The data span brightness temperatures of $\sim$1500--1800\,K, 1300--1600\,K, 1000--1300\,K, and 1200--1500\,K for the dayside, evening, nightside, and morning phases, respectively.
A behavior common to all phase-resolved spectra is that the brightness temperature of the planet steadily decreases with wavelength. These slopes in brightness temperature arise from the fact that the photospheric pressures gradually decrease at longer wavelengths, where temperatures are colder in the case of a non-inverted vertical thermal structure such as that of NGTS-10\,Ab. For the nightside, however, the rising brightness temperature upper limits at short wavelengths $<2\,\mu$m should not be strictly interpreted as increasing emission, but mainly as a gradual decrease in sensitivity to low brightness temperatures (Figure \ref{fig:nightside_wo_offset}).

We infer phase-resolved effective temperatures for NGTS-10\,Ab by computing the integrated $\lambda = 0.5$--$5.5$\,$\mu$m planetary flux from the spectra, propagating uncertainties in the data and system parameters ($R_\mathrm{p}/R_\star$ and $T_\mathrm{eff,\star}$), and determining the blackbody temperature that would produce the equivalent emission. This method yields effective temperatures of $T_\mathrm{day} = 1796\pm26$\,K, $T_\mathrm{evening} = 1560\pm27$\,K, $T_\mathrm{night} = 1180_{-82}^{+72}$\,K, and $T_\mathrm{morning} = 1349 \pm 34$\,K. These values are systematically lower (by $\sim$50\,K) than those of \citet{Parmentier2026}, with these differences likely arising from changes in the data reduction. Using the relations presented in \citet{Cowan_2011Ab}, we infer a Bond albedo of $A_\mathrm{B} = 0.03_{-0.11}^{+0.10}$ and a heat redistribution efficiency of $\varepsilon=0.380_{-0.073}^{+0.077}$ ($\varepsilon=0$ is no heat transport and $\varepsilon=1$ is full redistribution). These measurements are listed in Table \ref{table:phase_resolved_temps} of the Appendix, and we recommend they be considered instead of the values presented in \citet{Parmentier2026} as our larger uncertainty on $R_\mathrm{p}/R_\star$ results in more conservative estimates of NGTS-10\,Ab's energy budget.

To evaluate the robustness of our results to the treatment of the nightside spectrum, we also compute the nightside effective temperature from the non-shifted spectrum, considering only the data above 2\,$\mu$m where the flux is consistently positive. This yields a dayside effective temperature of $T_\mathrm{night} = 1042\pm26$\,K, consistent at $\sim$1.6$\sigma$ with the temperature inferred from the shifted spectrum. The lower effective temperature from the non-shifted spectrum is to be expected since the planet-to-star flux ratio is systematically lower by 250\,ppm, but also because the longer wavelengths ($>2\,\mu$m) overall probe lower pressures where temperatures are cooler. When repeating the computation of the Bond albedo and heat recirculation efficiency assuming $T_\mathrm{night} = 1042\pm26$\,K, we obtain values of $A_\mathrm{B} = 0.125_{-0.082}^{+0.076}$ and $\varepsilon = 0.254_{-0.024}^{+0.026}$. This corresponds to a higher Bond albedo and lower heat recirculation efficiency (by $0.7$ and $1.6\sigma$, respectively) since the cooler nightside temperature represents a lower total emitted energy and a greater day-night temperature gradient. Overall, this dependency of the nightside temperature measurement on the assumed spectrum does not impact our inference of a low Bond albedo and relatively poor heat redistribution efficiency for NGTS-10\,Ab.

The observed dayside spectrum of NGTS-10\,Ab is significantly hotter than the models that do not consider clouds (Fig. \ref{fig:FpFs_4_phases}), with reduced chi-squared values of $\chi^2_N$ = 20.4, 9.6, and 7.7 for the fiducial, $\tau_\mathrm{drag}$ = $10^6$\,s, and $\tau_\mathrm{drag}$ = $10^5$\,s models, respectively. While drag does lead to an increase in dayside temperature, the drag timescale value of $10^5$\,s is not sufficient to bring the model in agreement with the dayside spectrum. The cloudy models provide good fits to the dayside spectrum, producing $\chi^2_N$ values of 2.1 ($r_\mathrm{part}$ = 0.5\,$\mu$m) and 2.2 ($r_\mathrm{part}$ = 5\,$\mu$m). As for the nightside spectrum, it is broadly consistent with the fiducial model ($\chi^2_N$ = 1.4), the $r_\mathrm{part}$ = 5\,$\mu$m cloudy scenario ($\chi^2_N$ = 1.1), as well as the $\tau_\mathrm{drag}$ = $10^5$\,s ($\chi^2_N$ = 1.8) and $10^6$\,s models ($\chi^2_N$ = 1.8). Only the $r_\mathrm{part}$ = 0.5\,$\mu$m cloudy model provides an inadequate fit to the nightside ($\chi^2_N$ = 6.2), as it is considerably colder than the other GCMs due to the cloud opacity blocking the emission from the deeper layers of the atmosphere. The evening spectrum is best reproduced by the $r_\mathrm{part}$ = 5\,$\mu$m cloudy ($\chi^2_N$ = 2.7) and fiducial ($\chi^2_N$ = 3.9) models, whereas the cloudy $r_\mathrm{part}$ = 0.5\,$\mu$m scenario ($\chi^2_N$ = 12.1) as well as the $\tau_\mathrm{drag}$ = $10^5$\,s ($\chi^2_N$ = 18.6) and $\tau_\mathrm{drag}$ = $10^6$\,s ($\chi^2_N$ = 12) GCMs are systematically lower than the data. Finally, the morning phase is best fit by the cloudy $r_\mathrm{part}$ = $5\,\mu$m case ($\chi^2_N$ = 4.7) and the models including drag ($\chi^2_N$ = 3.6 and 5.9 for $\tau_\mathrm{drag}$ = $10^5$ and $10^6$\,s, respectively), while the fiducial ($\chi^2_N$ = 12.1) and $r_\mathrm{part}$ = $0.5\,\mu$m ($\chi^2_N$ = 15) GCMs are too cold to match the observations. Overall, the $r_\mathrm{part} = 5\,\mu$m model produces the lowest total reduced chi-square (averaged over the four phases), with a value of $\chi^2_{N\mathrm{,tot}}=2.7$.

If, instead of comparing the shifted nightside spectrum to the GCMs, we consider the non-shifted spectrum using only wavelengths above 2\,$\mu$m, we obtain reduced chi-square values of $\chi^2_N =$ 1.5, 3.4, 1.2, 1.1, and 1.1 for the fiducial, cloudy $r_\mathrm{part} = 0.5\,\mu$m, cloudy $r_\mathrm{part} = 5\,\mu$m, $\tau_\mathrm{drag} = 10^5$\,s, and $\tau_\mathrm{drag} = 10^6$\,s models, respectively. The treatment of the nightside spectrum therefore does not affect our conclusion that all models provide adequate fits to the nightside spectrum, except for the cloudy $r_\mathrm{part}=0.5\,\mu$m GCM which produces a nightside that is too cold and devoid of molecular features.

To assess the robustness of our conclusions to the choice of reduction, we also perform a comparison of the GCM spectra to those presented in \citet{Parmentier2026} (Fig. \ref{fig:FpFs_spectra_4_phases}). This comparison yields total reduced chi-square values of 17.2, 15.2, 4.2, 13.2, and 14.2 for the fiducial, $r_\mathrm{part} = 0.5\,\mu$m, $r_\mathrm{part} = 5\,\mu$m, $\tau_\mathrm{drag} = 10^5$\,s, and $\tau_\mathrm{drag} = 10^6$\,s models, respectively. The reduced chi-square values from the comparison with the \citet{Parmentier2026} spectra are generally higher due to the lower uncertainties, but they nevertheless support our conclusion that the $r_\mathrm{part} = 5\,\mu$m scenario best reproduces the observations.


For the nightside and morning spectra, we observe a notable discrepancy between the data and models over the 3--4\,$\mu$m range, where the planet's emission is significantly higher than predicted by the GCMs (Figure \ref{fig:FpFs_4_phases}). This discrepancy is caused by the depletion of methane in NGTS-10\,Ab's atmosphere, as demonstrated by \citet{Parmentier2026}. Equilibrium chemistry predicts methane to be the dominant carbon-bearing species at the low temperatures of the nightside and morning phases. However, advection of hot gas from the dayside to the nightside continually replenishes the colder regions in CO. This process of advection occurs on timescales that are much shorter than the CO-to-CH$_4$ conversion rate, resulting in a global atmospheric composition that is set by the dayside \citep[e.g.,][]{Agundez2014,Showman2020}. Although the models presented in this work were run assuming chemical equilibrium, future studies could explore the inclusion of disequilibrium chemistry in the GCMs \citep[e.g.,][]{Mehta2026} to reproduce the observed methane depletion.

\begin{figure*}
  \centering  \includegraphics[width=1.\textwidth]{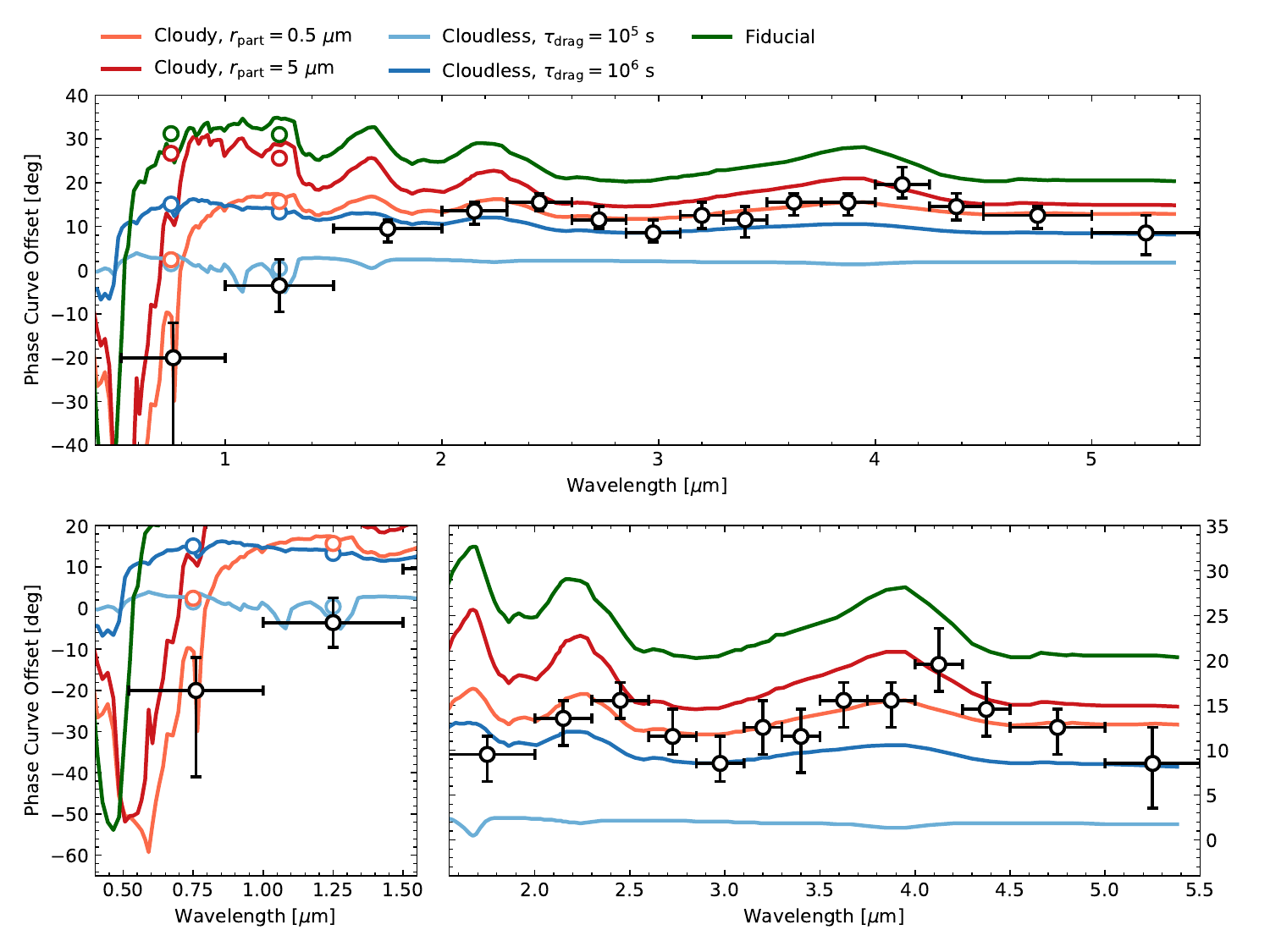}
  \vspace{-6mm}
  \caption{Spectroscopic phase curve offsets of NGTS-10\,Ab (black circles) as measured from the low-resolution light curves. The short-wavelength ($\lambda = 0.5$--$1\,\mu$m) offset is produced from the $\mathcal{R}_\mathrm{cloud}$-$T_\mathrm{cond}$ Lambertian fit, and the offsets above $1\,\mu$m are extracted from the second-order sinusoid fits. At short-wavelengths (lower left panel), the maximum of the phase curve occurs after secondary eclipse (westward offsets), whereas the maximum occurs before secondary eclipse in the infrared (lower right panel). Spectroscopic phase curve offsets from GCMs are shown for the fiducial (green), cloudy (orange and red), and atmospheric drag (light and dark blue) scenarios. The phase curve offset models, binned over the range of the two short-wavelength data points, are shown for comparison to the data. The moderate eastward offsets in the infrared can be explained either by nightside clouds or weak drag. Towards the optical, the cloudy models show partial cloud coverage on the western dayside and best reproduce the observed westward offset.}
  \label{fig:offset_vs_wave}
\end{figure*}

\subsection{Infrared Phase Curve Offsets}

At wavelengths greater than 1.5\,$\mu$m, where the planetary flux is dominated by thermal emission, we measure near-constant eastward offsets of $\sim$13$^\circ$ (Figure \ref{fig:offset_vs_wave}). In comparison, the GCM without clouds and atmospheric drag
produces infrared offsets that range from 20$^\circ$ to 33$^\circ$, two times larger than what is inferred from the data.

This discrepancy between the observed offsets and the values predicted by the fiducial GCM points towards the presence of an additional mechanism, such as clouds or atmospheric drag, that dampens the phase curve offsets. This is consistent with findings from previous works that cloudless and dragless models systematically overestimate the phase curve offsets \citep[e.g.,][]{Parmentier_2018,Parmentier_2020,Roth2024TiO}. We find that the phase curve offsets respond sharply to the inclusion of drag. Only the longest drag timescale explored in the GCM analysis ($\tau_\mathrm{drag} =10^6$\,s) is able to somewhat reproduce the observations, with offsets ranging from 5--13$^\circ$.
When reducing the drag timescale to $\tau_\mathrm{drag}=10^5$\,s, the offsets further decrease to 0--2$^\circ$ and are nearly constant at infrared wavelengths. Although a weak drag scenario can reproduce the observed phase curve offsets, it fails to match the observed dayside temperatures because a significant fraction of the absorbed heat is still transported to the nightside (Fig. \ref{fig:FpFs_4_phases}). Drag is therefore unable to self-consistently reproduce both the phase curve offsets and the phase-resolved spectra.

The inclusion of clouds in the GCMs also lead to a decrease of the thermal phase curve offsets.
The cloudy $r_\mathrm{part} = 0.5\,\mu$m and $r_\mathrm{part}=5\,\mu$m scenarios produce infrared offsets ranging from 11--17$^\circ$ and 14--25$^\circ$, respectively. Smaller cloud particle sizes lead to a larger day-night temperature gradient and sharper flux contrasts between the cloudy and cloudless regions of the atmosphere, resulting in slightly lower phase curve offsets. Additionally, nightside clouds inhibit radiative cooling on the nightside, leading to high dayside temperatures consistent with the observations.
Our phase curve offset measurements thus point to the presence of clouds, with no evidence for additional mechanisms impeding the winds on NGTS-10\,Ab.

\subsection{Optical Phase Curve Offsets}



Whereas the offsets are eastward (positive) in the infrared, they transition to westward (negative) towards the optical. At wavelengths below 1\,$\mu$m, the inclusion of clouds is needed to reproduce the observed westward phase curve offset (Fig. \ref{fig:offset_vs_wave}). Models with smaller cloud particles better fit the magnitude of the westward offset, as they result in a transition from westward to eastward offsets that occurs at longer wavelengths. The wavelength at which this transition occurs is directly related to the reflectivity of the clouds, which dictates the relative contribution of reflected light to the total planetary flux. Clouds with small grain sizes are more reflective (since particle number decreases and forward scattering increases with particle size; \citealt[e.g.,][]{Cuzzi2014,Parmentier2016}), and thus exhibit westward offsets over a larger wavelength range. However, despite being able to reproduce the trend of decreasing offsets with decreasing wavelengths, none of our cloudy models can quantitatively reproduce the offsets observed in the optical part of the PRISM bandpass. Smaller cloud particle sizes than those explored in the GCM could potentially reproduce the magnitude of the measured short-wavelength westward offset. 

The fiducial model also predicts westward offsets below 0.7\,$\mu$m. This asymmetry in reflected light is likely caused by the broadening of the sodium lines in the hottest regions of the atmosphere eastward of the substellar point, leading to an eastern dayside that is less reflective. This cannot account for our measured westward offset, however, as the fiducial model predicts an eastward offset (Fig. \ref{fig:offset_vs_wave}) and negligible planetary flux within the $\lambda=0.5$--$1\,\mu$m bandpass.

For the GCMs considering atmospheric drag, the dayside temperature distribution is more symmetric about the substellar point and the phase curve offsets are nearly constant at the shortest wavelengths, which does not match the observed trend of a transition to westward offsets towards the optical.

\subsection{Reflected Light and Thermal Maps}\label{sec:refl_therm_maps}

\begin{figure*}
  \centering  \includegraphics[width=0.9\textwidth]{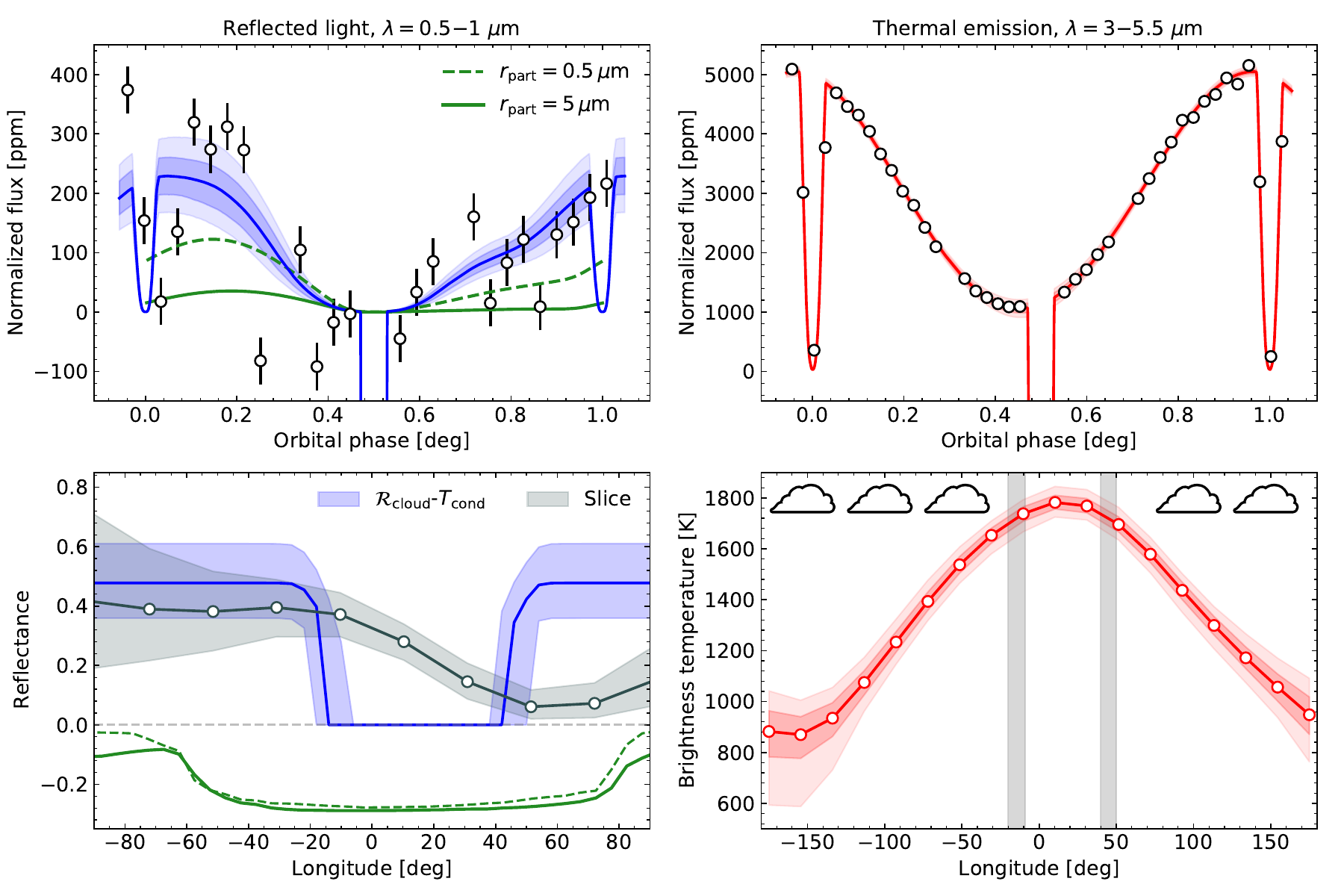}
  \caption{Constraints on the reflectance and temperature distributions of NGTS-10\,Ab from its broadband optical and infrared phase curves. Top left: Systematics-corrected broadband optical ($\lambda$ = 0.5--1\,$\mu$m) phase curve (black points), binned at intervals of 40 minutes (1,500 integrations per bin) for visual clarity. The median fit from the $\mathcal{R}_\mathrm{cloud}$--$T_\mathrm{cond}$ Lambertian model, along with the 1 and 2$\sigma$ confidence intervals are shown by the blue line and shaded regions, respectively. The reflected light curves from the $r_\mathrm{part}$ = 0.5\,$\mu$m (dashed green line) and 5\,$\mu$m (solid green line) cloudy GCMs, binned within the same bandpass as the data, are shown for comparison. Top right: Same as top left for the broadband infrared ($\lambda$ = 3--5.5\,$\mu$m) thermal phase curve, binned at intervals of 27 minutes (1,000 integrations per bin) for visual clarity. Bottom left: Median reflectance distributions from the $\mathcal{R}_\mathrm{cloud}$--$T_\mathrm{cond}$ Lambertian fit (blue line) and the slice fit (gray line), along with their 1$\sigma$ confidence intervals (shaded region), as measured from the optical phase curve. The transitions from cloudy to cloud-free and cloud-free to cloudy inferred from the $\mathcal{R}_\mathrm{cloud}$--$T_\mathrm{cond}$ fit occur at longitudes $\varphi$ = -17 to -6$^\circ$ and 37 to 48$^\circ$ (1$\sigma$ confidence), respectively. We show for comparison the latitude-averaged MgSiO$_3$ cloud tracer abundances for the $r_\mathrm{part}$ = 0.5\,$\mu$m (dashed green line; at 10$^{-3}$\,bar) and $r_\mathrm{part}$ = 5\,$\mu$m (solid green line; at 10$^{-2}$\,bar). Bottom right: Brightness temperature map distribution inferred from the infrared phase curve fit. The maximum of the distribution is shifted eastwards. The 1$\sigma$ constraints on the position of the cloudy-to-cloud-free transitions are indicated by the gray shaded regions.}
  \label{fig:refl_therm_maps}
\end{figure*}

From the thermal phase curve, we infer an infrared ($\lambda= 3$--$5.5\,\mu$m) offset of $13.3\pm1.2^\circ$, corresponding to a peak in emission before secondary eclipse (Figure \ref{fig:refl_therm_maps}), and a normalized phase curve amplitude ($[F_\mathrm{max}-F_\mathrm{min}]/F_\mathrm{max}$) of $A = 0.789\pm0.015$. We convert the longitudinal planetary flux ($F_\mathrm{p}/F_\star$) distribution to brightness temperatures by inverting the Planck function over the 3--5.5\,$\mu$m bandpass, accounting for the stellar spectrum and throughput of the NIRSpec PRISM instrument. We find that temperatures at the photospheric pressures of NGTS-10\,Ab change considerably with longitude, with a maximum in temperature of $\sim$1800\,K near the substellar point that gradually decreases to $\sim$900\,K around the antistellar point. 

As for the reflected light phase curve, we find that it is anti-correlated with the thermal phase curve. Indeed, the free slice fits yield a western dayside more reflective than the eastern dayside, as well as a temperature maximum that is shifted eastward of the substellar point (Fig. \ref{fig:refl_therm_maps}). This justifies the use of our $\mathcal{R}_\mathrm{cloud}$--$T_\mathrm{cond}$ model, where we assume the cloud and temperature distributions are intertwined, as a physically-motivated scenario. When fitting for the cloud reflectance and condensation temperature, we infer values of $\mathcal{R}_\mathrm{cloud}= 0.47_{-0.12}^{+0.14}$ and $T_\mathrm{cond} = 1721_{-21}^{+18}$\,K for the Lambertian kernel. Above the inferred condensation temperature, the atmosphere is cloud-free, resulting in an eastern dayside that is devoid of clouds and non-reflective for longitudes ranging from approximately -17 to 48$^\circ$ (Figure \ref{fig:refl_therm_maps}). When assuming an isotropic kernel, we constrain a cloud reflectance and a condensation temperature of $\mathcal{R}_\mathrm{cloud} = 0.882_{-0.081}^{+0.063}$ and $T_\mathrm{cond} = 1735_{-21}^{+13}$\,K, respectively. The Lambertian and isotropic kernels produce phase curves that are broadly consistent in shape and amplitude (Figure \ref{fig:refl_light_fits}), resulting in constraints on the condensation temperature that are consistent within 1$\sigma$. The main difference in results between the two fits is the inferred cloud reflectance, which is to be expected since the scaling between reflectance and geometric albedo for the isotropic kernel strongly deviates from the linear relationship produced by Lambertian reflection \citep[e.g.,][]{Madhusudhan2012}. 

For the reflected light phase-curve fits using the slice method, we observe that the inferred distribution strongly depends on the assumed regularization term $\gamma_R$ (Figure \ref{fig:RL_maps_vs_gamma}). If $\gamma_R$ is kept free, the fit prefers strong penalization against second derivatives and produces a dayside reflectance map that linearly decreases towards the eastern dayside. We repeat the analysis by fixing $\gamma_R$ to values of 0.1, 1, and 10, and find that $\gamma_R = 1$ is the regularization strength that provides the greatest trade-off between map flexibility and smoothness of the resulting solution. The $\gamma_R = 1$ map is in qualitative agreement with the distribution inferred from the $\mathcal{R}_\mathrm{cloud}$--$T_\mathrm{cond}$ fit (Figure \ref{fig:refl_therm_maps}), with both methods inferring a western dayside that is more reflective than the eastern dayside. The exact shape of the reflectance distribution depends on the model, however, as the slice technique cannot reproduce the sharp transitions from cloudy to cloud-free that are modeled with the $\mathcal{R}_\mathrm{cloud}$--$T_\mathrm{cond}$ fit because of the low slice resolution of 10$^\circ$ and the penalization against large second derivatives. The $\mathcal{R}_\mathrm{cloud}$--$T_\mathrm{cond}$ method is also itself limited, since the range of possible position and spacing of the cloudy to cloud-free transitions is dictated by the brightness temperature map. Indeed, because the temperature distribution used for the $\mathcal{R}_\mathrm{cloud}$--$T_\mathrm{cond}$ fit is shifted eastward, this method can only produce cloud distributions that are uniform or shifted westward. Both methods thus have their individual strengths and weaknesses, and provide a complementary view of NGTS-10\,Ab's reflectance distribution. 

Because of the relatively low SNR of the reflected light phase curve, whose fit is further complicated by the presence of the two exposure breaks, there are important variations in the inferred phase curve offset depending on the model that is considered. The double-sinusoid and shifted Lambertian models produce offsets of $-31.6_{-4.0}^{+5.0}\,^{\circ}$ and $-12.0_{-7.6}^{+8.4}\,{^\circ}$, respectively. As for the $\mathcal{R}_\mathrm{cloud}$--$T_\mathrm{cond}$ fits, they result in phase curve offsets of $-20_{-21}^{+8}\,{^\circ}$ for the Lambertian kernel and $-8.6_{-2.9}^{+2.7}\,{^\circ}$ for the isotropic kernel. Finally, the slice method produces an offset of $-17.5_{-5.8}^{+6.6}\,^{\circ}$ when fixing the regularization strength to $\gamma_R = 1$. This spread in phase curve offsets caused by the choice of parameterization highlights the difficulty of reflected light measurements, as the signal amplitude is significantly lower than that of infrared phase curves. We note that the retrieved phase curve properties could also potentially depend on the choice of systematics model, as long term trends in the data can induce correlations with the astrophysical signal of interest. When comparing the phase curve parametrizations, we find that the $\mathcal{R}_\mathrm{cloud}$--$T_\mathrm{cond}$ Lambertian model produces the lowest chi-square, with a value of $\Delta \chi^2 = -14$ relative to the second-order sinusoid fit (Fig. \ref{fig:refl_light_fits}). Nevertheless, all models yield virtually the same fit quality, with reduced chi-square values of $\chi^2_N \approx 1.26$. Therefore, no model can be considered preferred above the others based on this metric. A common trend among all parameterizations, however, is that they show negative phase curve offsets $>1\sigma$ from 0,
resulting from a non-uniform reflectance distribution where the western dayside is more reflective than the eastern dayside.



\begin{figure*}
  \centering  \includegraphics[width=1.0\textwidth]{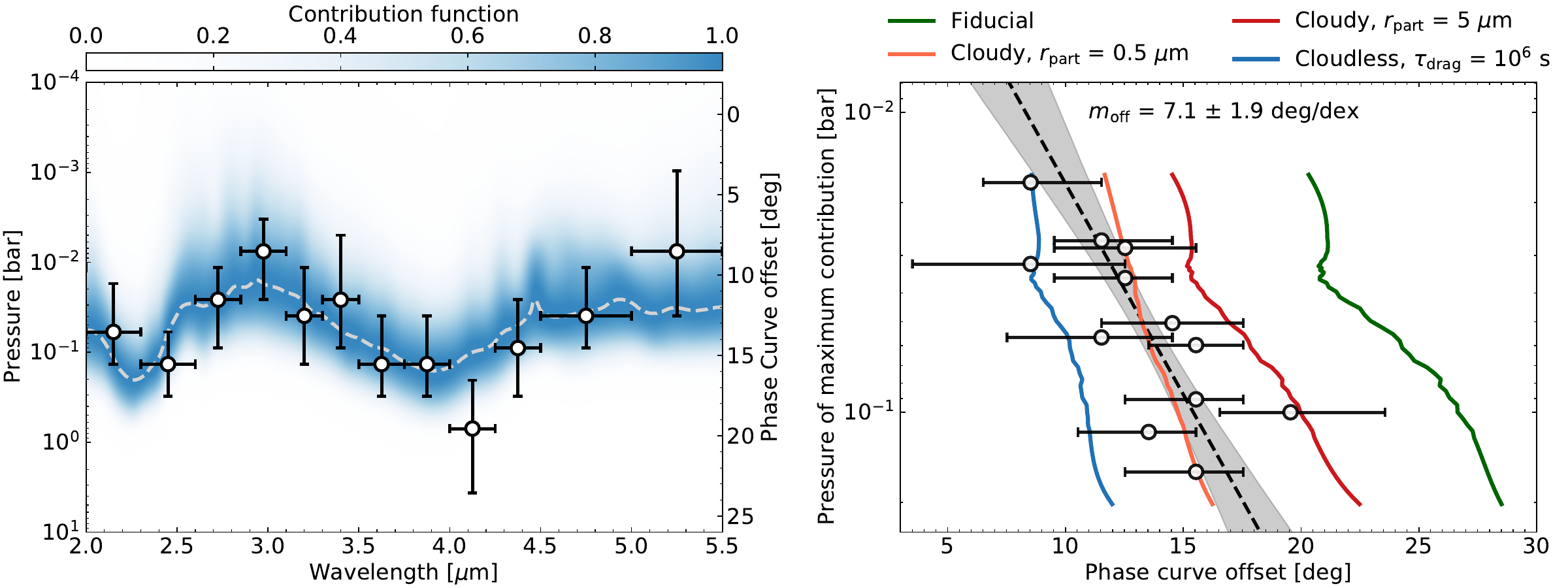}
  \caption{Altitude-dependent heat redistribution in the atmosphere of NGTS-10\,Ab. Left: Phase-curve offsets of NGTS-10\,Ab (black circles) at infrared wavelengths ($\lambda\geq2\,\mu$m). Color represents the normalized contribution of the pressure levels to the outgoing flux at a given wavelength, as computed from the ScCHIMERA dayside spectrum best-fit model presented in \citet{Parmentier2026}. The gray dashed line represents the pressure of maximum contribution as a function of wavelength. The scale of the y-axis for the phase-curve offsets was manually adjusted to match the overall amplitude of the contribution function. Right: Phase-curve offsets (black points) as a function of their pressure of maximum contribution. Fitting a linear trend to the offsets as a function of $\log p$ yields a slope of $m_\mathrm{off} = 7.1\pm1.9$\,degrees per pressure dex, in broad agreement with the trends predicted by GCMs.}
  \label{fig:offs_vs_press}
\end{figure*}

\subsection{Altitude-Dependent Offsets}\label{sec:offsets_vs_pressure_analysis}

At infrared wavelengths ($\lambda \gtrsim1.5\,\mu$m), the GCMs predict spectral variations in the phase curve offsets that directly correlate with the presence of spectral features (mainly H$_2$O and CO; Fig. \ref{fig:offs_vs_press}). These variations arise from the different altitudes that are probed in and out of molecular bands \citep{Dobbs-Dixon2017}. The radiative and dynamical timescales of the atmosphere can change considerably with pressure, leading to modulations in the efficiency of heat transport. Atmospheric drag strongly decreases the amplitude of molecular features in the spectroscopic offsets, as made evident by the $\tau_\mathrm{drag}=10^5$\,s GCM which is virtually constant with wavelength. Smaller cloud particle sizes also reduce the amplitude of the offsets' modulation, although to a lesser extent than drag. To evaluate whether the variations in our infrared offsets coincide with spectral features, we compare them to the 
dayside contribution function of NGTS-10\,Ab (Fig. \ref{fig:offs_vs_press}). 
The magnitude of the offsets appears to follow the shape of the H$_2$O and CO features covered over the 2--5.5\,$\mu$m wavelength range. We measure lower phase curve offsets in the H$_2$O and CO bands compared to the continuum, indicative of poorer heat redistribution efficiency at high altitudes. 

We can further map the spectroscopic offsets to specific pressure levels using the contribution function (Fig. \ref{fig:offs_vs_press}). The altitude probed by an offset is assumed to correspond to the pressure of maximum contribution (the pressure at which $\mathcal{C}$ is highest) given the wavelength at which it was observed. 
Following \citet{Stevenson2014}, we parameterize the relation between offsets and pressure as a linear trend ($\mathrm{offset} = m_\mathrm{off}\log p_\mathrm{max} + b_\mathrm{off}$) and fit for the slope $m_\mathrm{off}$ and intercept $b_\mathrm{off}$ using the \texttt{curve\_fit} function of \texttt{scipy.optimize}. The fit is performed considering all wavelengths $\lambda>2\,\mu$m, discarding the shorter wavelengths to avoid potential contamination from reflected light signal. We find a positive correlation between offsets and pressure, and infer a slope of $m_\mathrm{off}=7.1\pm1.9$\,deg per pressure dex,
3.8$\sigma$ away from the null hypothesis (no correlation, $m_\mathrm{off}=0$). By fitting a linear trend to the offset--pressure curves from the GCMs, we measure slopes of 9.0\,deg\,dex$^{-1}$ for the fiducial model, as well as 4.1 and 7.9\,deg\,dex$^{-1}$ for the cloudy models with $r_\mathrm{part} = 0.5$ and 5\,$\mu$m, respectively. The models with drag also show lower slopes than the fiducial model, with values of -0.1\,deg\,dex$^{-1}$ for $\tau_\mathrm{drag} = 10^5$\,s and 3.5\,deg\,dex$^{-1}$ for $\tau_\mathrm{drag} = 10^6$\,s. Except for the $\tau_\mathrm{drag} = 10^5$\,s model, all GCMs produce offset--pressure slopes within $2\sigma$ of the measured value. Finally, we repeat our offset--pressure analysis using the phase curve offsets from the analysis presented in \citet{Parmentier2026} and infer a slope of $m_\mathrm{off} = 9.2\pm2.5$ (3.7$\sigma$ above 0), in agreement at 0.7$\sigma$ with our measurement.

\section{Discussion}\label{sec:discussion}

Our observations of the hot-Jupiter NGTS-10\,Ab demonstrate the vast diagnostic power of spectroscopic exoplanet phase curves, revealing a climate that is shaped by heat transport, a heterogeneous cloud distribution, and heat redistribution efficiency that varies significantly with altitude. We discuss in the sections below the mechanisms that dictate NGTS-10\,Ab's atmosphere, and extrapolate our findings to the population of hot- and ultra-hot Jupiters with optical and infrared phase curve measurements.

\subsection{Distinguishing Atmospheric Drag and Nightside Clouds}


In the case of a cloudless atmosphere with no atmospheric drag, 
the predicted infrared phase curve offsets
are significantly greater than our measurements. This finding indicates that there is a mechanism at play reducing the phase curve offset, either through reduced eastward advection due to atmospheric drag or the presence of nightside clouds. The overestimation of phase curve offsets by fiducial GCMs has been a common trend in hot-Jupiter observations \citep[e.g.,][]{Parmentier_2018,Parmentier_2020,Roth2024TiO}. However, limited photometric precision and wavelength coverage have previously prevented the inhomogeneous clouds and atmospheric drag scenarios from being clearly distinguished. Moreover, we find that the cloudless model is unable to reproduce the overall shape of the phase-resolved spectra, with the largest discrepancy observed for the dayside and morning spectra. This disagreement between model and data results from the significant eastward advection of heat towards the nightside, which effectively cools the dayside and evening regions and results in lower emission at these phases compared to the observations. 


An effect able to reduce the thermal phase curve offset and increase the day-night contrast is that of atmospheric drag. This process
can be attributed to a variety of phenomena such as magnetic drag, turbulence, and instabilities which decrease wind speeds \citep[e.g.,][]{Goodman2009,Perna2010,Fromang2016} and, consequently, the efficiency of heat redistribution. When drag is considered in the models, we find that the long-wavelength phase curve offsets decrease rapidly with decreasing drag timescales.
Although a drag timescale intermediate to $\infty$ and $10^6$\,s could reproduce the magnitude of the observed offsets, it would produce dayside and evening spectra that are too dim to match the data (Figure \ref{fig:FpFs_4_phases}). Even for drag timescales of $10^5$\,s, at which point the phase curve offsets are virtually null, the dayside temperature is not sufficiently high to match the observations. 

This raises the question: how can the phase curve offset be null if there remains significant heat redistribution to the nightside? 
This effect is elucidated by \citet{roth2024a}, which demonstrates that when including weak-to-intermediate drag in GCMs, it is the eastward jet that is first impeded by drag forces. The circulation regime of the atmosphere then transitions from one that is dominated by eastward heat transport to one dominated by winds moving radially away from the substellar point towards the nightside. In the latter regime, the phase curve offset is low since the temperature distribution is approximately symmetric about the substellar point. Nevertheless, the substellar-to-antistellar symmetric flow is able to sustain heat advection to the nightside, and the day-to-night contrast gradually increases as drag timescales decrease. It is thus apparent that, in contrast with past modeling work \citep[e.g.,][]{Cowan2011,Zhang2017}, the phase curve offset and amplitude cannot be considered as two perfectly coupled quantities. Rather, the full shape of the thermal phase curve is needed to properly distinguish between competing atmospheric scenarios.


An inhomogeneous cloud distribution on NGTS-10\,Ab provides the most likely explanation for its low phase curve offsets and large day-night contrast. In this scenario, there is notable eastward transport of heat and the nightside is covered in clouds as local temperatures are sufficiently low for condensation to occur. 
As the heat from the dayside reaches the nightside, the upwards flux is effectively blocked at pressures below the cloud deck, and the heat instead pursues its eastward trajectory back to the dayside. This mechanism is therefore able to efficiently increase the dayside temperature while also conserving non-zero phase curve offsets, in contrast with drag processes which inevitably dampen eastward advection. The muted offsets from the cloudy GCMs do not result from longer advection timescales, but rather from the sharp longitudinal gradients in flux that are introduced by the clouds \citep{Parmentier_2020}. Through the convolution from spatial heat distribution to phase curve, these gradients create the illusion that the offset of the thermal distribution is closer to the substellar point than it is in reality. We find that both models including clouds
broadly reproduce the measured dayside spectrum. For the evening and nightside spectra, however, the model with $r_\mathrm{part} = 5\,\mu$m better matches the observations, as smaller particle sizes result in low emission and muted features at these phases. The cloudy GCMs are also in overall agreement with the measured phase curve offsets, with the 0.5\,$\mu$m model best reproducing the exact level of the infrared offsets.


\subsection{Altitude-dependent Dynamics}

To first order, the ability of an atmosphere to redistribute heat from the dayside to the nightside is set by the radiative ($\tau_\mathrm{rad}$) and advective ($\tau_\mathrm{adv}$) timescales \citep{Showman2002}. The radiative timescale is related to the efficiency of the atmosphere to re-radiate heat back towards space, and can be approximated as 

\begin{equation}
    \tau_\mathrm{rad} \sim \frac{p}{g}\frac{c_p}{4\sigma T^3},
\end{equation}

\medskip 
\noindent 
where $p$ is the pressure, $g$ the surface gravity, $c_{p}$ the specific heat capacity, and $T$ the temperature. The hotter a parcel of gas is, the faster it cools. Moreover, radiative timescales vary as a function of pressure within the atmosphere. At low pressures, heat is re-radiated faster than in the deep atmosphere. This is consistent with the findings of \citet{Iro2005}, who found that radiative timescales increase monotonically with pressure in their 1D time-dependent radiative modeling of HD 209458\,b. As for the advective timescale, it is commonly defined as the time needed for a parcel of gas to be advected over a distance of one planetary radius for a given wind speed $v_\mathrm{wind}$

\begin{equation}
    \tau_\mathrm{adv} \sim \frac{R_\mathrm{p}}{v_\mathrm{wind}}.
\end{equation}

\medskip  
The heat redistribution efficiency and thermal phase curve offset are related to the ratio of these two timescales ($\xi=\tau_\mathrm{rad}/\tau_\mathrm{adv}$). If $\xi\ll1$, advection is inefficient since the atmosphere re-radiates heat too quickly for it to be transported and, conversely, $\xi\gg1$ corresponds to a scenario where advection is efficient. Because the ratio $\xi$ depends on the pressure, temperature, and wind speed, which all vary with altitude, we expect the heat redistribution efficiency of the atmosphere to change as different pressure levels are probed. From the $r_\mathrm{part} = 5\,\mu$m cloudy GCM, we find that the equatorial zonal-mean wind speed increases from 4.9 to 6\,km\,s$^{-1}$ between 1 and 0.01\,bar (Figure \ref{fig:zonal_mean_wind}), corresponding to a $\sim$20\% variation in $\tau_\mathrm{adv}$. As for the dayside-averaged temperature, it decreases from 1600 to 1900\,K over the same pressure range, leading to a 40\% reduction in $\tau_\mathrm{rad}$. Comparatively, the pressure decreases by two orders of magnitude, producing a factor of one hundred variation in the radiative timescale. As pressure decreases, the atmosphere rapidly becomes more efficient at radiating heat away, leading to a reduced advection efficiency. This is consistent with our finding of a negative offset--altitude correlation, and is also supported by the GCMs which show a decreasing longitude of maximum hemisphere-averaged temperature with decreasing pressure (Fig. \ref{fig:GCM_maps}). It is therefore apparent that, at the photospheric pressures of hot-Jupiters, the variation of heat transport efficiency with altitude is dominated by changes in the radiative timescales, primarily through variations in pressure, rather than by changes in wind speeds.





Altitude-dependent advection has also been observed in WASP-43\,b from its HST/WFC3 phase curve \citep{Stevenson2014}. The offsets were found to be systematically lower inside the 1.4\,$\mu$m water absorption feature compared to the out-of-band wavelengths.
When repeating the same analysis described in Section \ref{sec:offsets_vs_pressure_analysis} on the WASP-43\,b HST offsets (discarding the same outlier wavelengths as \citealt{Stevenson2014}), we measure an offset--pressure slope of $m_\mathrm{off}$ = 18.0 $\pm$ 3.5\,deg\,dex$^{-1}$ (Figure \ref{fig:offs_vs_pressure_W43b}). The amplitude of the inferred slope is 5.1$\sigma$ away from 0, consistent with \citet{Stevenson2014} who measured a correlation significance of 5.6$\sigma$. While NGTS-10\,Ab and WASP-43\,b both show the same behaviour of decreasing offsets with altitude, the amplitude of the slope is $\sim$2.5 times larger for WASP-43\,b, possibly indicative of a difference in the dynamical regimes of these two planets despite the similarity of their physical properties. Observations of WASP-43\,b with NIRSpec G395H (JWST GTO 1224, P.I.: S. Birkmann), which cover a similar wavelength range as our observations of NGTS-10\,Ab, will enable a more robust comparison to our offset measurements.

\subsection{The Interplay of Temperature and Clouds}

The temperature and cloud structures of an atmosphere are closely intertwined: cloud formation is set by the local vertical thermal structure, and clouds in turn impact the thermal structure by reflecting a portion of the incoming stellar flux and blocking the upwards planetary flux. This results in a fine interplay between clouds and thermal structure in hot-Jupiter atmospheres. Indeed, as shown in Figure \ref{fig:refl_therm_maps}, we find the reflectance and brightness temperature distributions of NGTS-10\,Ab to be anti-correlated. The brightest point of the atmosphere at infrared wavelengths is located to the east of the substellar point, and the optical phase curve is best explained by a western hemisphere that is cloudier than the eastern hemisphere.

A common assumption is that winds are the main driver of both the cloud and temperature distributions. Winds transport heat eastwards from the substellar point, and could also transport clouds from the nightside towards the western dayside, explaining the observed reflected light and thermal emission asymmetries. This assumption is likely incorrect, however, as clouds transported from the nightside to the dayside would be out of equilibrium, and evaporation timescales are predicted to be significantly shorter ($\tau_\mathrm{evap} \approx 10^{-1}$--$10^3$\,s for $\mu$m-sized MgSiO$_3$ grains; \citealt{Powell2018}) than the advection timescales ($\tau_\mathrm{adv}\approx 10^4$--$10^5$\,s). Advection is therefore unable to replenish the dayside in clouds at a rate sufficient to counter evaporation. Instead, the probable explanation is that winds primarily shape the temperature distribution, which then dictates the cloud distribution. As heat is transported eastward, the western dayside becomes considerably colder than the eastern dayside. This difference in temperature is responsible for the preferential formation of clouds west of the substellar point. Beyond a certain longitude over the dayside, temperatures become sufficiently elevated for clouds to evaporate, leading to an eastern dayside that is mostly cloud-free. As the circulation pattern of hot-Jupiters evolves with varying equilibrium temperature, orbital period, and atmospheric composition, the cloud distribution is thus expected to closely follow variations in temperature distributions.

In addition to phase curves, the limb-asymmetry technique \citep[e.g.,][]{Jones2020}, where the difference in apparent size between the morning (west) and evening (east) terminators is measured through the shape of a transit's ingress and egress signal, provides complementary information about exoplanetary climates. A common trend from recent JWST observations is that hot-Jupiters tend to show morning terminators that are colder and cloudier than the evening terminators \citep[e.g.,][]{Espinoza2024,Fu2025,Mukherjee_2026}. Notably, the hot-Jupiter WASP-94\,Ab ($T_\mathrm{eq} \approx 1500$\,K), which has an equilibrium temperature close to that of NGTS-10\,Ab, shows a morning terminator dominated by cloud opacity along with a relatively clear evening \citep{Mukherjee_2026}. When computing the temperature difference between the evening ($\varphi = 90^\circ$) and morning ($\varphi = -90^\circ$) terminators from our inferred brightness temperature map (Fig. \ref{fig:refl_therm_maps}), we measure a value of $\Delta T_\mathrm{terminators}=203_{-62}^{+57}\,$K. Although we cannot directly compare evening--morning temperature gradients measured from phase curves and transits, as they probe distinct pressure levels in the atmosphere, our measurement is broadly consistent within uncertainties with the gradient of $449\pm83$\,K measured for WASP-94\,Ab.
It is therefore apparent that the interplay of cloud and temperature observed from our NGTS-10\,Ab phase curve likely also shapes the limb-resolved spectra obtained with the JWST.

\begin{figure}
  \centering  \includegraphics[width=0.48\textwidth]{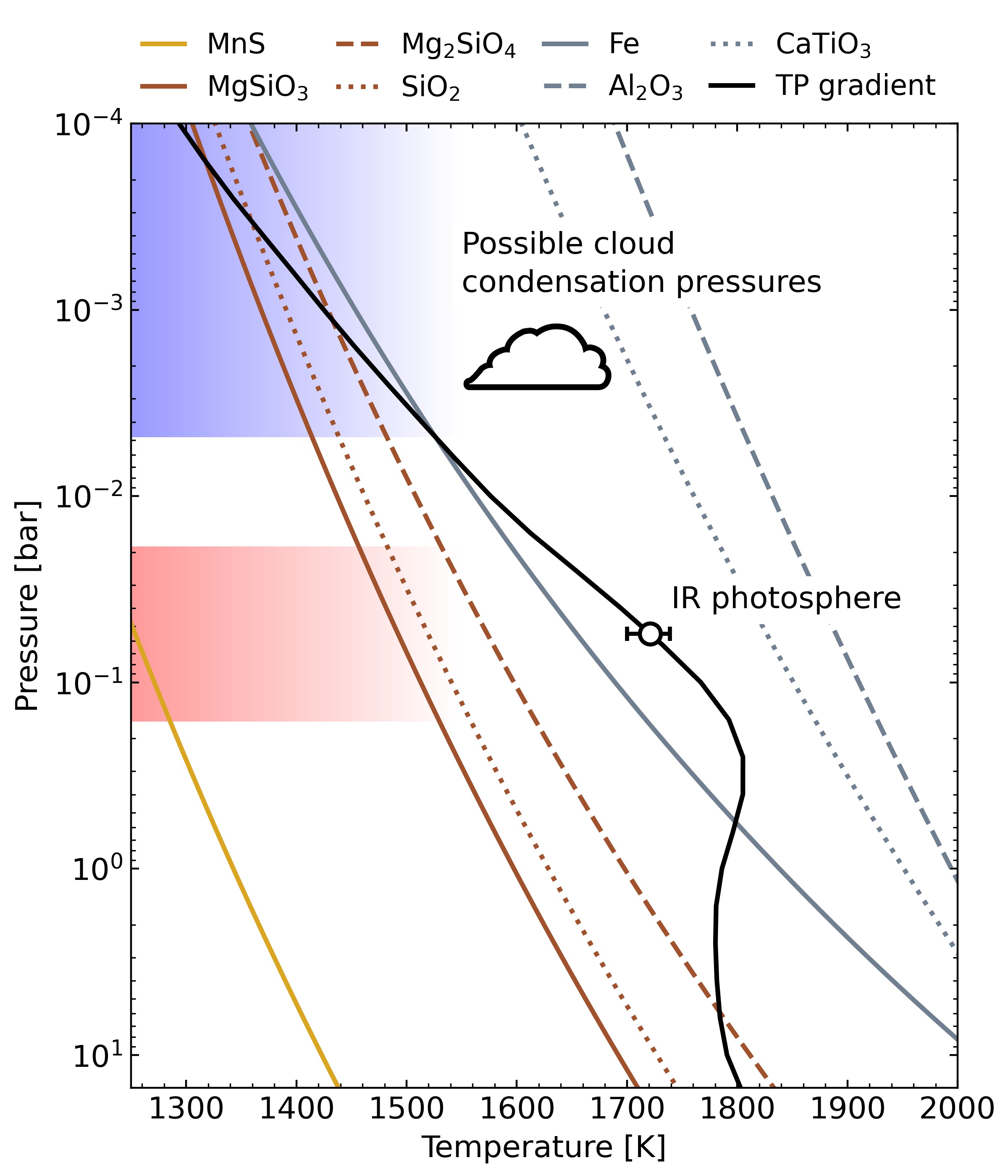}
  \vspace{-6mm}
  \caption{Measured condensation temperature $T_\mathrm{cond}$ from the reflected light phase curve fit (black point). We assume that the condensation temperature is probed at pressures $p = 20$--$150$\,mbar following the 3--5.5\,$\mu$m contribution function from Figure \ref{fig:offs_vs_press}. We compare the measured condensation temperature to condensation curves for sulfide (MnS; gold), silicate (MgSiO$_3$, Mg$_2$SiO$_4$, and SiO$_2$; brown), and metal (Fe, CaTiO$_3$, and Al$_2$O$_3$; gray) clouds \citep{Visscher2010,Morley2012,Wakeford2017,Grant2023}. The inferred condensation is extrapolated to other pressures using the gradient of the TP-profile measured from the best-fit ScCHIMERA dayside retrieval (black line), which crosses the condensation curves of silicate and iron clouds at low pressures ($p = 0.15$--$5$\,mbar).}
  \label{fig:Tcond_vs_cond_curves}
\end{figure}

\subsection{Cloud Properties of NGTS-10\,Ab}



We compare in Figure \ref{fig:Tcond_vs_cond_curves} our inferred condensation temperature ($T_\mathrm{cond} = 1721_{-21}^{+18}$\,K) from the $\mathcal{R}_\mathrm{cloud}$--$T_\mathrm{cond}$ Lambertian fit to condensation curves of various sulfide, silicate, and metal cloud species. At first glance, the inferred condensation temperature does not match with any of the considered cloud species. This is because $T_\mathrm{cond}$ is measured at the pressures of the photosphere ($p=20$--$150$\,mbar) since it is derived from the shape of the 3--5.5\,$\mu$m thermal map. However, the clouds probed through reflected light do not necessarily condense at photospheric pressures, and could instead form at higher altitudes. Therefore, our $T_\mathrm{cond}$ measurement should be interpreted as a measurement of the photospheric temperature at the longitudes where the cloudy-to-cloudless transitions occur, rather than as an absolute condensation temperature.
To circumvent this limitation, we extrapolate our photospheric condensation temperature measurement to other altitudes using the gradient of the dayside temperature-pressure profile (taken from the best-fit ScCHIMERA retrieval; \citealt{Parmentier2026}). This yields an inferred cloud condensation temperature curve $T_\mathrm{cond}(p) = T(p) + [T_\mathrm{cond,phot} - T(p_\mathrm{phot})]$, where $T(p)$ is the temperature-pressure profile, $T_\mathrm{cond,phot}$ our measured condensation temperature, and $T(p_\mathrm{phot})$ the photospheric temperature of the temperature-pressure profile. We can then compare our $T_\mathrm{cond}(p)$ profile with the cloud species' condensation curves over the full pressure range of the atmosphere. 
As shown in Fig. \ref{fig:Tcond_vs_cond_curves}, our condensation temperature profile crosses the condensation curves of silicate (MgSiO$_3$, Mg$_2$SiO$_4$, SiO$_2$) and iron clouds at lower pressures ($p = 0.15$--$5$\,mbar).
This indicates that the observed clouds are likely one of or a combination of these species. It is unlikely, however, that iron clouds are responsible for the observed reflected light signal since they are significantly less reflective than silicate clouds for most particle sizes \citep[e.g.,][]{Morris_2024}.

The preference for large grain sizes from the phase-resolved spectra (Fig. \ref{fig:FpFs_4_phases}) is similar to the results from the MIRI LRS phase curve of WASP-43\,b. Indeed, nightside clouds on WASP-43\,b were inferred from the shape of its broadband phase curve, whose amplitude and offset is best reproduced by cloudy GCMs compared to cloudless models \citep{bell2024nightsidecloudsdisequilibriumchemistry}. However, its nightside spectrum notably did not reveal any evidence of cloud spectral features, such as the 9\,$\mu$m silicate feature commonly observed in brown dwarfs \citep[e.g.,][]{Cushing2006}. Because the amplitude of the silicate feature is inversely proportional to the cloud particle size, any potential clouds on the planet's nightside must therefore be composed of relatively large ($\gtrsim1\,\mu$m) grains. Nevertheless, we note that no GCMs of WASP-43\,b including the effect of atmospheric drag were compared to the observations in \citet{bell2024nightsidecloudsdisequilibriumchemistry}, preventing a definitive conclusion as to the source of WASP-43\,b's large day-night contrast and low phase curve offset.

\subsection{Need for Increased Cloud Complexity in GCMs}

An apparent discrepancy appears when comparing the measured phase-resolved spectra and the reflected light phase curve to the cloudy GCMs. A large cloud particle size is needed to properly fit the phase-resolved spectra, but the amplitude of the optical phase curve is best reproduced by clouds with small grain sizes. Indeed, the $r_\mathrm{part} = 0.5\,\mu$m and $5\,\mu$m GCMs produce reflected light signals that are approximately 2 and 6 times weaker than the measured phase curve, respectively (Figure \ref{fig:refl_therm_maps}). This difference in amplitude likely arises from the assumption of a fixed particle size distribution in the GCMs, whereas microphysical cloud modeling studies have found that clouds in hot-Jupiter atmospheres may host bimodal size distributions with distinct cloud bases and vertical extents \citep{Powell2018,Powell2024}. Under the scenario of a bimodal cloud particle size distribution, the thermal phase curve could be set by the large particles, with opacities that are approximately constant with wavelength. As for the reflected light phase curve, its large amplitude could result from a distinct population of small particles ($r_\mathrm{part}<0.5\,\mu$m) that efficiently reflect the incoming stellar irradiation at short wavelengths. For small particles, the opacity follows to first order a $\lambda^{-4}$ relation (Rayleigh scattering regime), leading to a negligible impact on the thermal phase curve. The existence of a population of small particles could also potentially resolve the discrepancy between the inferred longitudinal cloud distribution and that of the GCMs (Figure \ref{fig:refl_therm_maps}), as the spatial extent might spread over a larger portion of the dayside for small grain sizes. It is thus apparent that further cloud modeling complexity, such as coupled cloud microphysics and 3D atmospheric dynamics simulations, is possibly needed to reconcile the distinct cloud regimes that are probed from optical and infrared observations.

\begin{figure*}
  \centering  \includegraphics[width=1.\textwidth]{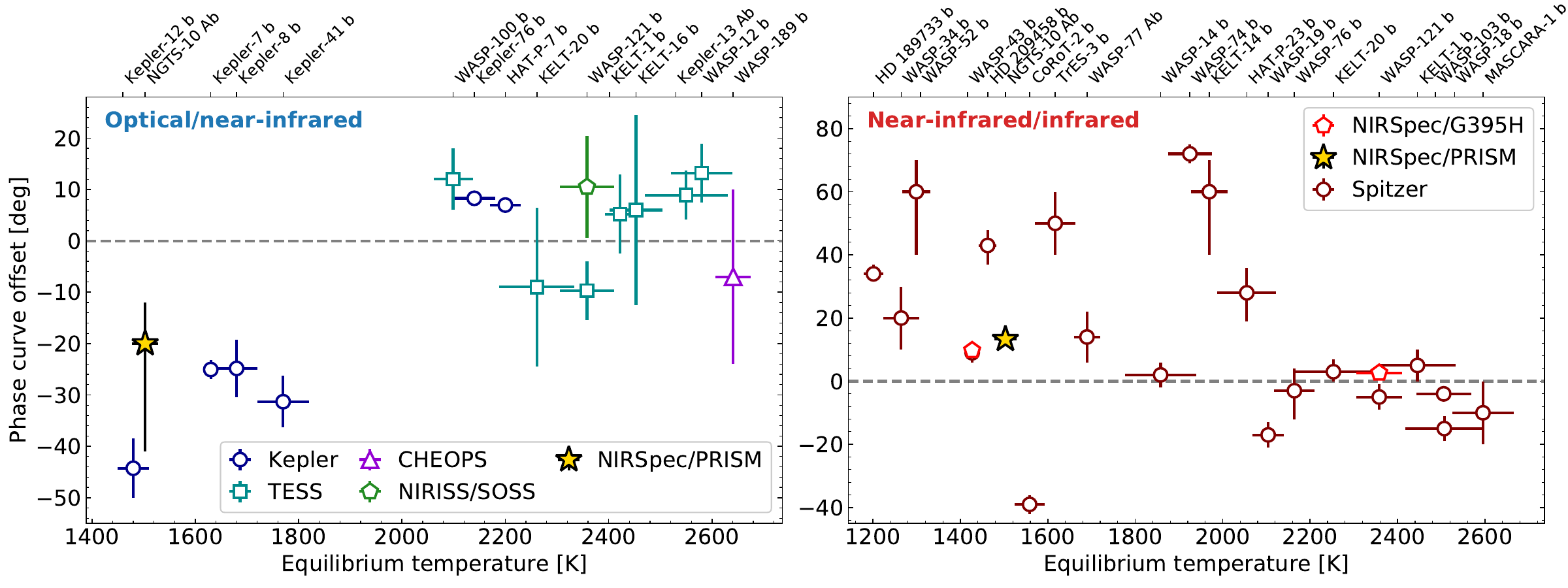}
  \caption{Phase curve offsets of hot- and ultra-hot Jupiters as a function of equilibrium temperature, measured at optical/near-infrared wavelengths (left) and near-infrared/infrared wavelengths (right). Phase curve offsets at optical wavelengths are generally westward (negative) and go to zero at high equilibrium temperatures, whereas offsets in the infrared are typically eastward (positive) at low temperatures and gradually decrease with temperature. 
  The references from which the phase curve offsets were taken are listed in Section \ref{sec:offset_refs} of the Appendix.}
  \label{fig:pop_phase_curve_offsets}
\end{figure*}

\subsection{NGTS-10\,Ab in the Context of the Hot-Jupiter Population}

When comparing the measured reflected light and thermal emission phase curve offsets of NGTS-10\,Ab to that of past measurements obtained with Kepler, CHEOPS, TESS, Spitzer, and JWST, we find that it fits well within the general trends observed for highly-irradiated gas giants (Fig. \ref{fig:pop_phase_curve_offsets}). We use the reflected light offset obtained from the $\mathcal{R}_\mathrm{cloud}$--$T_\mathrm{cond}$ Lambertian fit for this comparison.
Hot-Jupiters, especially in the $T_\mathrm{eq} =$ 1400--2000\,K regime, commonly show westward offsets at optical wavelengths and eastward offsets in the infrared. However, the lack of planets observed in both regimes has prevented a definitive statement on whether these westward and eastward offsets result from a common mechanism, or whether arise from distinct effects. Our finding of both a westward offset ($-20_{-21}^{+8}\,{^\circ}$) in the optical and an eastward offset ($13.3\pm1.2^\circ$) in the infrared for NGTS-10\,Ab, resulting from the interaction between heat transport and cloud coverage, unites these two patterns and provides evidence for a common cause to the trends observed from past reflected light and thermal emission measurements.

Additionally, optical and infrared measurements both trend towards low phase curve offsets at higher equilibrium temperatures. For thermal phase curves, this phenomenon can be attributed to shorter radiative timescales and the advent of strong drag forces which impede the flow of heat to the nightside \cite[e.g.,][]{Arcangeli2019,Coulombe2023}. Hotter objects thus tend to have more symmetric temperature distributions about their substellar point, meaning that any potential cloud distribution on their dayside would also be symmetric and would produce a reflected light phase curve that peaks at mid-eclipse. Furthermore, for the hottest objects, the contribution of thermal emission remains significant, and can even dominate, at optical and near-infrared wavelengths. This contamination of thermal emission at short wavelengths complicates the interpretation of photometric measurements such as those obtained with TESS and CHEOPS \citep[e.g.,][]{Bourrier2020,Deline2022}.

While the trends of decreasing phase curve offsets at increasing temperatures appear robust both in reflected light and thermal emission, it is unclear whether the scatter in thermal offset measurements at temperatures $T_\mathrm{eq}\lesssim2100$\,K arises from an intrinsic diversity in the properties of hot-Jupiter atmospheres, or whether it is caused by the important systematics that affected Spitzer phase curve observations. For example, WASP-43\,b and NGTS-10\,Ab, which have both been observed with JWST, show relatively low phase curve offsets compared to other planets observed in the same temperature range (e.g., WASP-52\,b, HD 209458\,b, TrES-3\,b).
Further optical and infrared phase curve observations of hot-Jupiters at the population level are needed to isolate the contribution of parameters such as orbital period, temperature, and atmospheric composition to the circulation and cloud coverage of highly-irradiated gas giants. Thus far, JWST phase curve observations have focused primarily on the two ends of the hot/ultra-hot Jupiter population, covering three planets in the $T_\mathrm{eq}$ $\approx$ 1200--1500\,K range (HD 189733\,b, WASP-43\,b, and NGTS-10\,Ab) and five in the $T_\mathrm{eq}\gtrsim2200$\,K regime (WASP-76\,b, WASP-121\,b, WASP-103\,b, WASP-12\,b, and KELT-9\,b). Notably, the $T_\mathrm{eq} = $1600--2200\,K region of the parameter space remains poorly explored in both thermal emission and reflected light. Important effects such as changes in the dominant cloud species (e.g., from silicates to metal oxides), the onset of thermal dissociation, and the transition from non-inverted to inverted thermal structures are predicted to occur over this temperature range, making it a key region of interest for future exploration. 


\section{Conclusion}\label{sec:conc}

In this work, we present an analysis of the red-optical to mid-infrared JWST/NIRSpec PRISM phase curve of the hot-Jupiter NGTS-10\,Ab. Leveraging the 1.4\,s precision on the planet's mid-transit time, we refine constraints on its possible orbital decay and find it to be consistent with 0 ($\dot{P} = -4.6\pm4.2$\,ms\,yr$^{-1}$). The spectroscopic phase curve ($\lambda = 0.5$--$5.5\,\mu$m), which covers almost the full SED of NGTS-10\,Ab, yields phase-resolved effective temperatures of $T_\mathrm{day}=1796\pm26$\,K, $T_\mathrm{evening} = 1560\pm27$\,K, $T_\mathrm{night} = 1180_\mathrm{-82}^{+72}$\,K, and $T_\mathrm{morning} = 1349\pm34$\,K. These measurements are indicative of a low Bond albedo for the planet ($A_\mathrm{B} = 0.03_{-0.11}^{+0.10}$) and relatively inefficient heat redistribution ($\varepsilon=0.380_{-0.073}^{+0.077}$).

The spectroscopic phase curve offsets reveal the existence of two regimes for NGTS-10\,Ab's atmospheric pattern, with westward offsets at short wavelengths ($\lambda<1\,\mu$m) and eastward offsets in the infrared. Although both clouds and atmospheric drag can explain the observed dampened infrared phase curve offsets compared to a cloudless and dragless model, comparison of the measured phase-resolved spectra to GCMs strongly rules out the drag scenario. Instead, the observations are best explained by an atmosphere with clouds covering the nightside and a portion of the western dayside. This is further confirmed by our retrieved temperature and reflectance maps from the broadband infrared ($\lambda=3$--$5.5$\,$\mu$m) and optical/near-infrared ($\lambda=0.5$--$1.0\,\mu$m) phase curves of NGTS-10\,Ab. The two maps are anti-correlated, with the eastern dayside being devoid of clouds at the longitudes where temperatures reach their peak.

Additionally, modulation in the infrared offsets is found to be correlated with the presence of molecular features. The phase curve offsets are systematically lower inside molecular bands, yielding a slope between offsets and pressure of $m_\mathrm{off} = 7.1\pm1.9$ degrees per pressure dex in the 1--0.01\,bar range. This decrease in heat transport efficiency at higher altitudes arises primarily from variations in the radiative timescale, which scales directly with pressure. 

The simultaneous reflected light and thermal emission coverage enabled by the JWST/NIRSpec PRISM data bridges for the first time hot-Jupiter phase curve measurements of westward offsets in the optical and eastward offsets in the infrared, unveiling a common trend within this population that arises from the interplay between heat circulation and cloud coverage. Furthermore, we demonstrate the diagnostic power of spectroscopic phase curves in breaking degeneracies between competing atmospheric scenarios, such as inhomogeneous cloud cover and atmospheric drag, which commonly complicate the interpretation of atmospheric characterization from secondary eclipses alone.

Additional observations of NGTS-10\,Ab in reflected light and thermal emission could provide further constraints on the properties and spatial extent of its clouds. Notably, NGTS-10\,Ab lies in the first PLATO field \citep{Rauer2025}, which will be monitored continuously for two continuous years at high cadence \citep{Nascimbeni2025}. These observations of the planet over the 0.5--1\,$\mu$m bandpass will refine its reflected light phase curve measurement. Complementary thermal emission observations with JWST/MIRI LRS would also provide additional information on the composition and particle size of NGTS-10\,Ab's clouds through the 9\,$\mu$m silicate feature.

More broadly, the growing sample of visible-to-infrared phase curves from the JWST and the upcoming Ariel mission \citep{Charnay2022_ariel} will enable a population-level view of dynamical transport and its coupling to cloud formation in highly irradiated gas giants. The $T_\mathrm{eq} = 1600$--$2200$\,K regime stands out as a key target for future phase curve characterization, as it remains poorly explored in reflected light and thermal emission and is expected to exhibit important transitions in chemistry, dominant cloud species, and thermal structure. The insights gained from these worlds will inform our understanding of cloud formation, circulation, and climate regimes across a wide range of planetary conditions, including smaller and cooler planets for which phase curve studies are currently unachievable, and will establish hot-Jupiters as important analogs to young, low-gravity brown dwarfs.




\section*{Acknowledgements}
\noindent 
We thank the anonymous reviewer for their thorough review and insightful comments that improved the quality of this work.
This work is based on observations made with the NASA/ESA/CSA James Webb Space Telescope. The data were obtained from the Mikulski Archive for Space Telescopes at the Space Telescope Science Institute, which is operated by the Association of Universities for Research in Astronomy, Inc., under NASA contract NAS 5-03127 for JWST. These observations are associated with program \#2158. The specific observations analyzed can be accessed via \dataset[DOI: 10.17909/v6je-q140]{https://doi.org/10.17909/v6je-q140}. L.-P.C. thanks Jonathan Gagné, Björn Benneke, and Loïc Albert for useful discussions and feedback. He also thanks Nishil Mehta for help with GCM post-processing, and Amélie Gressier for valuable suggestions on the figures. L.-P.C. acknowledges financial support from Mitacs through the Mitacs Accelerate program, in partnership with the Montreal Planetarium, and from the Canadian Space Agency through grants 25JWGO4B01 and 25JWGO4B03. X.T. is supported by the National Natural Science Foundation of China (grant No. 42475131). J.M.D acknowledges support from the research program VIDI New Frontiers in Exoplanetary Climatology with project number 614.001.601, which is (partly) financed by the Dutch Research Council (NWO).

\medskip
\noindent 
\textit{Software:} \texttt{Astropy} \citep{Astropy2013,Astropy2018,Astropy2022}, \texttt{batman} \citep{Kreidberg_2015}, \texttt{emcee} \citep{Foreman_Mackey_2013}, \texttt{Eureka!} \citep{Bell2022}, \texttt{Matplotlib} \citep{Hunter2007}, \texttt{NumPy} \citep{Harris2020}, and \texttt{SciPy} \citep{Virtanen2020}.

\clearpage
\appendix
\twocolumngrid

\renewcommand{\thefigure}{A\arabic{figure}}
\renewcommand{\theHfigure}{A\arabic{figure}}

\renewcommand{\thetable}{A\arabic{table}}
\renewcommand{\theHtable}{A\arabic{table}}
\setcounter{figure}{0}
\setcounter{table}{0}



\section{Orbital Decay Analysis}\label{sec:orb_decay_analysis}

For our orbital decay analysis, we consider two scenarios: a static orbit (no decay) and a orbit that decays at a constant rate. In the absence of orbital decay, the planet's mid-transit time $T_n$ at a given epoch $n$ can be described as a linear relation

\begin{equation}
    T_n = T_\mathrm{mid} + nP,
\end{equation}

\medskip
\noindent 
where $T_\mathrm{mid}$ is the transit time at a reference epoch $n=0$, and $P$ is the orbital period. If, instead, the period decreases with time following a constant decay rate $\dot{P}$, the transit time would follow a quadratic relation \citep[e.g.,][]{Patra2017,Yee_2019}

\begin{equation}
    T_n = T_\mathrm{mid} + nP + \frac{n^2}{2}P\dot{P}.
\end{equation}

\medskip  
Recently, \citet{Griffiths2026} presented a transit-timing analysis of NGTS-10\,Ab considering 12 mid-transit time measurements (four transits initially presented in \citealt{McCormac_2020}, and eight new Danish Telescope/TESS transits) taken over eight years. When considering orbital decay, they infer a decay rate of $\dot{P} = -11.5\pm5.4$\,ms\,yr$^{-1}$, consistent with zero at $2.1\sigma$. We repeat this analysis considering the transit times of \citet{Griffiths2026} and our measured JWST/NIRSpec transit time. 

We perform the linear transit time analysis by fitting for the mid-transit time ($T_\mathrm{mid}$, $\mathcal{U}[15.0815,15.0835]$\,BJD-2460000) and orbital period ($P$, $\mathcal{U}[0.7668,0.7670]$\,days). We explore the parameter space with \texttt{emcee}, using 100 walkers and iterating for 10,000 steps, and report the posterior distributions from the last 4,000 steps. We measure a transit time of $T_\mathrm{mid} = 2460015.082430 \pm 1.6\times10^{-5}$\,BJD$_\mathrm{TDB}$ and an orbital period of $P = 0.766893299\pm 4.5\times10^{-8}$\,days. We repeat the same procedure for the quadratic transit time analysis, this time including the orbital decay rate ($\dot{P}$, $\mathcal{U}[-1000,1000]$ day\,day$^{-1}$) in the fit, and obtain a transit time $T_\mathrm{mid} = 2460015.082432\pm1.6\times10^{-5}$\,BJD$_\mathrm{TDB}$, period $P=0.76689315\pm1.5\times10^{-7}$\,days, and decay rate $\dot{P} = -4.6\pm4.2$\,ms\,yr$^{-1}$. As a sanity check, we repeat the quadratic analysis without the NIRSpec PRISM transit time and measure a decay rate of $\dot{P} = -11.5\pm5.4$ ms\,yr$^{-1}$, in perfect agreement with the measurement of \citet{Griffiths2026}.

\begin{figure*}
    \centering
    \includegraphics[width=1.\linewidth]{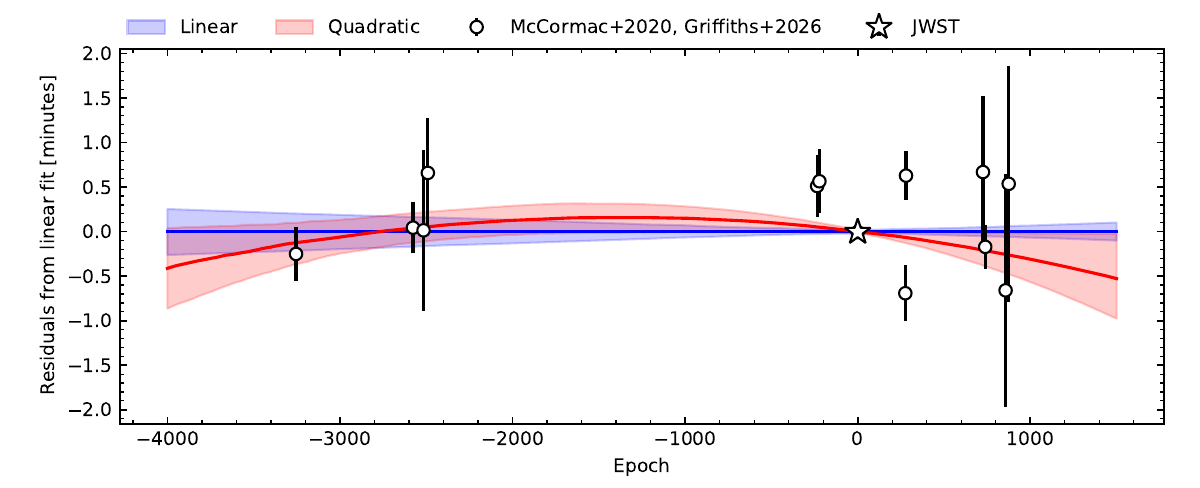}
    \caption{Residuals from a linear fit to the 13 transit time measurements of NGTS-10\,Ab. The median and 1$\sigma$ confidence intervals from  the linear and quadratic fits are shown in blue and red, respectively. The measured curvature from the quadratic fit is consistent with zero. The JWST transit time measurement dominates the fit near Epoch 0 due to its comparatively small uncertainty.}
    \label{fig:dPdt_fit}
\end{figure*}

The inclusion of the JWST transit time further constrains NGTS-10\,Ab's orbital decay rate to $\dot{P}=-4.6\pm4.2$\,ms\,yr$^{-1}$, in agreement with the results from \citet{Griffiths2026} and consistent with zero orbital decay at $1.1\sigma$. We show the residuals from the linear and quadratic analyses in Figure \ref{fig:dPdt_fit}.
The fits are dominated by the JWST transit time, which is 10--60 times more precise than the other measurements, and the residuals from the linear fit show no evidence for curvature. This non-detection of a decay rate is in line with the predictions from \citet{Barker2020} that rapid orbital decay is not expected given NGTS-10\,A's age of 10\,Gyr, unless gravity waves are fully damped. It is also in agreement with \citet{Tokuno2024}, who inferred a large stellar tidal quality factor ($Q_\star'\gtrsim10^8$) and predicted negligible transit timing shifts over the coming decades.
Additional high-SNR transit observations of NGTS-10\,Ab over a longer baseline will constrain its orbital decay rate at higher precision and shed further light unto its fate.




\section{Reflected Light Phase Curve Modeling}\label{sec:refl_light_models}

We describe in the sections below the various reflected light phase curve parameterizations applied to the short-wavelength observations of NGTS-10\,Ab.

\subsection{Shifted Lambertian}

We perform fits of 
the $\lambda = 0.5-1\,\mu$m broadband phase curve
using a shifted Lambert sphere model \citep[e.g.,][]{Esteves2015}, which has been commonly used to model Kepler reflected light phase curves. For this parameterization, the phase curve shape is set by the flux amplitude $F$ and the phase offset $\delta$:

\begin{equation}
\begin{split}
F_\mathrm{p}(t) = \frac{F}{\pi} \left[\sin z + (\pi- z )\cos z\right]~~~~~~~\\
\cos z = \sin i \cos\left[-2\pi\left(\frac{T_\mathrm{sec}-t}{P} - \delta\right)\right],
\end{split}
\end{equation}

\medskip 
\noindent
where $i$ is the inclination. We repeat the same light-curve fitting analysis as the one performed using the second-order sinusoid model, with the sole difference being that we only consider two parameters, $F$ ($\mathcal{U}[0,10000]$\,ppm) and $\delta$ ($\mathcal{U}[-0.5,0.5]$), for the phase curve model.

\subsection{Isotropic Kernel $\mathcal{R}_\mathrm{cloud}$--$T_\mathrm{cond}$ Fit}

Although we test both the $\mathcal{R}_\mathrm{cloud}$--$T_\mathrm{cond}$ parameterization and the slice method to assess the impact of the model on the retrieved reflectance distribution, these two fits assume the same underlying phase function. To address this, we also perform a fit using the $\mathcal{R}_\mathrm{cloud}$--$T_\mathrm{cond}$ parameterization where we instead consider an isotropic kernel

\begin{equation}
\mathcal{K}_{\mathrm{refl},n} = \frac{R_\mathrm{p}}{a\pi} \int_{\Omega_n} \frac{VI}{V+I}H(V,\mathcal{R})H(I,\mathcal{R})\mathrm{d}\Omega,
\end{equation}

\medskip 
\noindent 
where $H(\mu,\mathcal{R})$ is Chandrasekhar's H-function \citep{Chandrasekhar1960}, which depends on the viewing/illumination angle ($\mu$) and the reflectance $\mathcal{R}$. Rather than solving the H-function iteratively, we use the approximate form of \citet{Hapke1993}

\begin{multline}
H(\mu,\mathcal{R}) = \bigg[1 - (1-\sqrt{1-\mathcal{R}})\mu\cdot \\
\left(r_0 + \left[1-\frac{r_0}{2} - r_0\mu\right]\ln\frac{1+\mu}{\mu}\right)\bigg]^{-1},
\end{multline}

\medskip 
\noindent 
where $r_0 = \left[2/(1 + \sqrt{1-\mathcal{R}})\right] - 1$. This form is accurate to $\sim$1$\%$, which is sufficient given the precision of our observations. Using this kernel, we repeat the analysis described in Section \ref{sec:refl_light_mods}.




\section{Collection of Hot-Jupiter Phase Curve Offsets}\label{sec:offset_refs}

To compare with our measured reflected light and thermal emission phase curve offsets of NGTS-10\,Ab, we collate past optical, near-infrared, and infrared measurements from the literature. We describe and reference these measurements in the sections below.

\subsection{Optical/near-infrared offsets}

For the optical and near-infrared offsets, we consider photometric phase curve measurements made with the Kepler (0.4--0.9\,$\mu$m), TESS (0.6--1.0\,$\mu$m), and CHEOPS (0.35--1.1\,$\mu$m) space telescopes, as well as offsets measured from the short-wavelength coverage of the NIRISS/SOSS (0.6--2.8\,$\mu$m) and NIRSpec/PRISM (0.5--5.5\,$\mu$m) instruments of the JWST. 

The Kepler offsets are taken from \citet{Esteves2015} for Kepler-7\,b ($-25.1\pm1.9^\circ$), Kepler-8\,b ($-24.8\pm5.6^\circ$), Kepler-12\,b ($-44.3\pm5.8^\circ$), Kepler-41\,b ($-31.3\pm5.0^\circ$), Kepler-76\,b ($8.3\pm1.2^\circ$), and HAT-P-7\,b ($6.97\pm0.29^\circ$).

For the TESS phase curves, we take the offset measurements from \citet{Wong2020} and \citet{Wong_2021} for planets WASP-121\,b ($-9.7\pm5.7^\circ$), WASP-100\,b ($12.0{_{-5.7}^{+6.3}}^\circ$), KELT-1\,b ($5.2{_{-7.4}^{+8.0}}^\circ$), KELT-16\,b ($6{_{-19}^{+18}}^\circ$), KELT-20\,b ($-9{_{-15}^{+16}}^\circ$), Kepler-13\,Ab ($8.9{_{-4.6}^{+5.0}}^\circ$), and WASP-12\,b ($13.2\pm5.7^\circ$).

For CHEOPS, we consider the measured offset of WASP-189\,b ($-7\pm17^\circ$; \citealt{Deline2022}). 

Finally, we also include the offset inferred from the NIRISS SOSS order 2 ($\lambda = 0.6$--$0.85\,\mu$m) white light phase curve of WASP-121\,b ($10.5\pm9.9^\circ$; \citealt{Splinter2025}).




\subsection{Infrared offsets}

For the thermal phase curve offsets, we consider photometric phase curve measurements made with Spitzer and JWST covering the $\lambda = 3.5$--$5.5\,\mu$m wavelength range.

The Spitzer offsets are taken from the $4.5\,\mu$m photometric phase curves analyzed in \citet{Dang2025}, only considering measurements with $<20^\circ$ precision. The values considered are: HD\,189733\,b ($34{_{-2}^{+3}}^\circ$), WASP-34\,b ($20\pm10^\circ$), WASP-52\,b ($60{_{-20}^{+10}}^\circ$), WASP-43\,b ($9\pm3^\circ$), HD 209458\,b ($43{_{-6}^{+5}}^\circ$), CoRoT-2\,b ($-39\pm3^\circ$), TrES-3\,b ($50\pm10^\circ$), WASP-77\,Ab ($14\pm8^\circ$), WASP-14\,b ($2\pm4^\circ$), WASP-74\,b ($72\pm3^\circ$), KELT-14\,b ($60{_{-20}^{+10}}^\circ$), HAT-P-23\,b ($28{_{-9}^{+8}}^\circ$), WASP-19\,b ($-17\pm4^\circ$), WASP-76\,b ($-3{_{-9}^{+7}}^\circ$), KELT-20\,b ($3{_{-3}^{+4}}^\circ$), WASP-121\,b ($-5\pm4^\circ$), KELT-1\,b ($5\pm5^\circ$), WASP-18\,b ($-4\pm2^\circ$), WASP-103\,b ($-15\pm4^\circ$), and MASCARA-1\,b ($-10\pm10^\circ$).

We also include offset measurements made with the JWST/NIRSpec G395H NRS2 instrument, which covers a wavelength range similar to that of the Spitzer $4.5\,\mu$m photometric filter. Two hot-Jupiters currently have available NIRSpec G395H offset measurements: WASP-43\,b ($9.8\pm0.9^\circ$; \citealt{Challener2024}) and WASP-121\,b ($2.66\pm0.12^\circ$; \citealt{Evans2023}).

\clearpage

\begin{table}[hbt!]
\caption{Constraints on the energy budget of NGTS-10\,Ab. The effective temperature of the dayside, nightside, morning, and evening phases are computed from the phase-resolved spectra ($\lambda = 0.5$--$5.5\,\mu$m) which cover the bulk of the planetary SED. The Bond albedo ($A_\mathrm{B}$) and heat recirculation efficiency ($\varepsilon$) are computed from the dayside and nightside temperatures following \citet{Cowan_2011Ab}. The temperature gradient between the evening and morning terminators ($\Delta T_\mathrm{terminators}$) is inferred from the $\lambda = 3$--$5.5\,\mu$m thermal phase curve fit.}
\vspace{3mm}
\centering
\begin{tabular}{lcc}
\hline
\hline
Parameter     & Measurement         \\
\hline
$T_\mathrm{eff,day}$ [K]    &  $1796\pm26$ \\
$T_\mathrm{eff,night}$ [K] & $1180_{-82}^{+72}$ \\
$T_\mathrm{eff,morning}$ [K] & $1349\pm34$ \\
$T_\mathrm{eff,evening}$ [K] & $1560\pm27$ \\
$A_\mathrm{B}$  & $0.03_{-0.11}^{+0.10}$ \\
$\varepsilon$  & $0.380_{-0.073}^{+0.077}$ \\
$\Delta T_\mathrm{terminators}$ [K] & $203_{-62}^{+57}$ \\
\hline
\label{table:phase_resolved_temps}
\end{tabular}
\end{table}

\begin{figure}
    \centering
    \includegraphics[width=1.\linewidth]{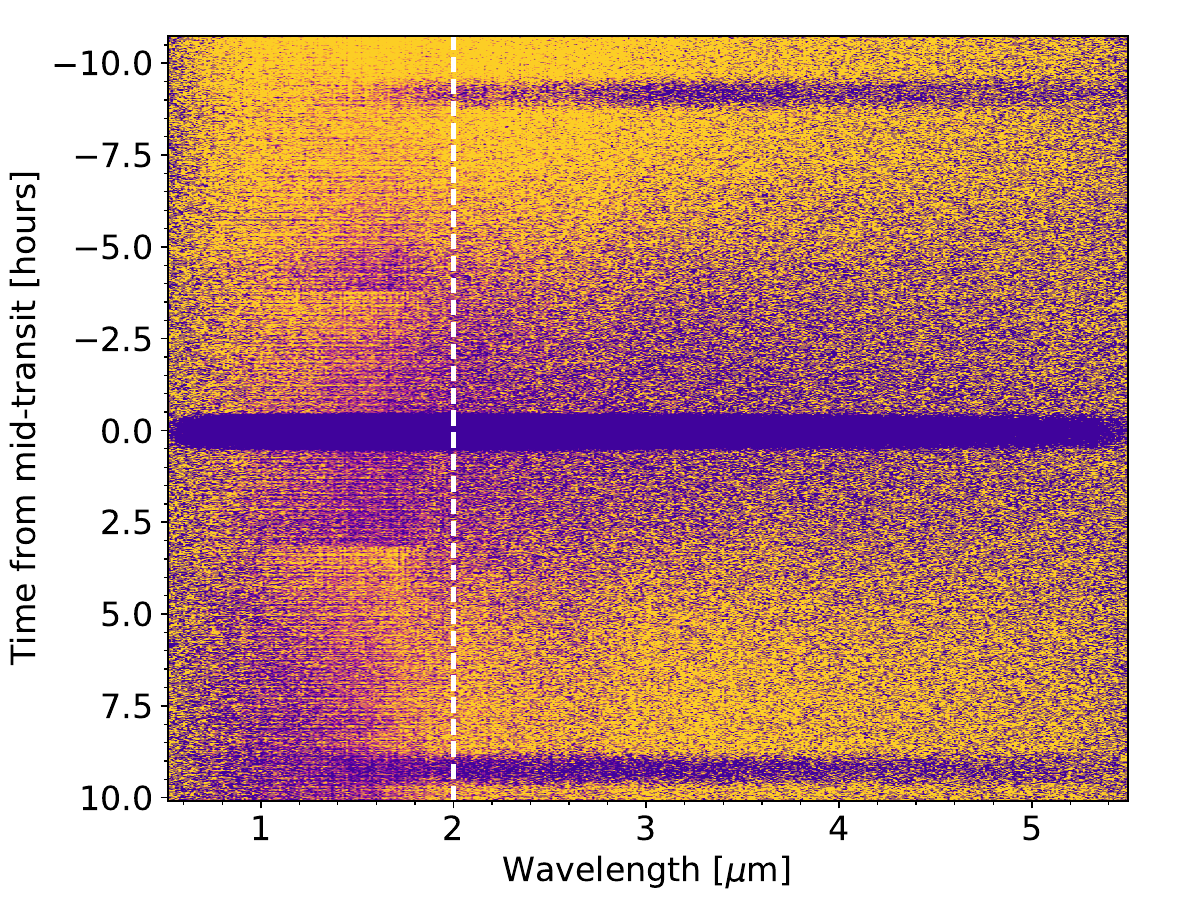}
    \vspace{-4mm}
    \caption{Raw spectroscopic light curves of NGTS-10\,Ab, shown at the pixel resolution of the instrument. The secondary eclipses are visible at the top and bottom of the figure, with the primary transit in between. A white dashed line is plotted at 2\,$\mu$m, below which the systematics present in the light curves are more pronounced.}
    \label{fig:2D_timeseries}
\end{figure}

\begin{figure}
    \centering
    \includegraphics[width=1.\linewidth]{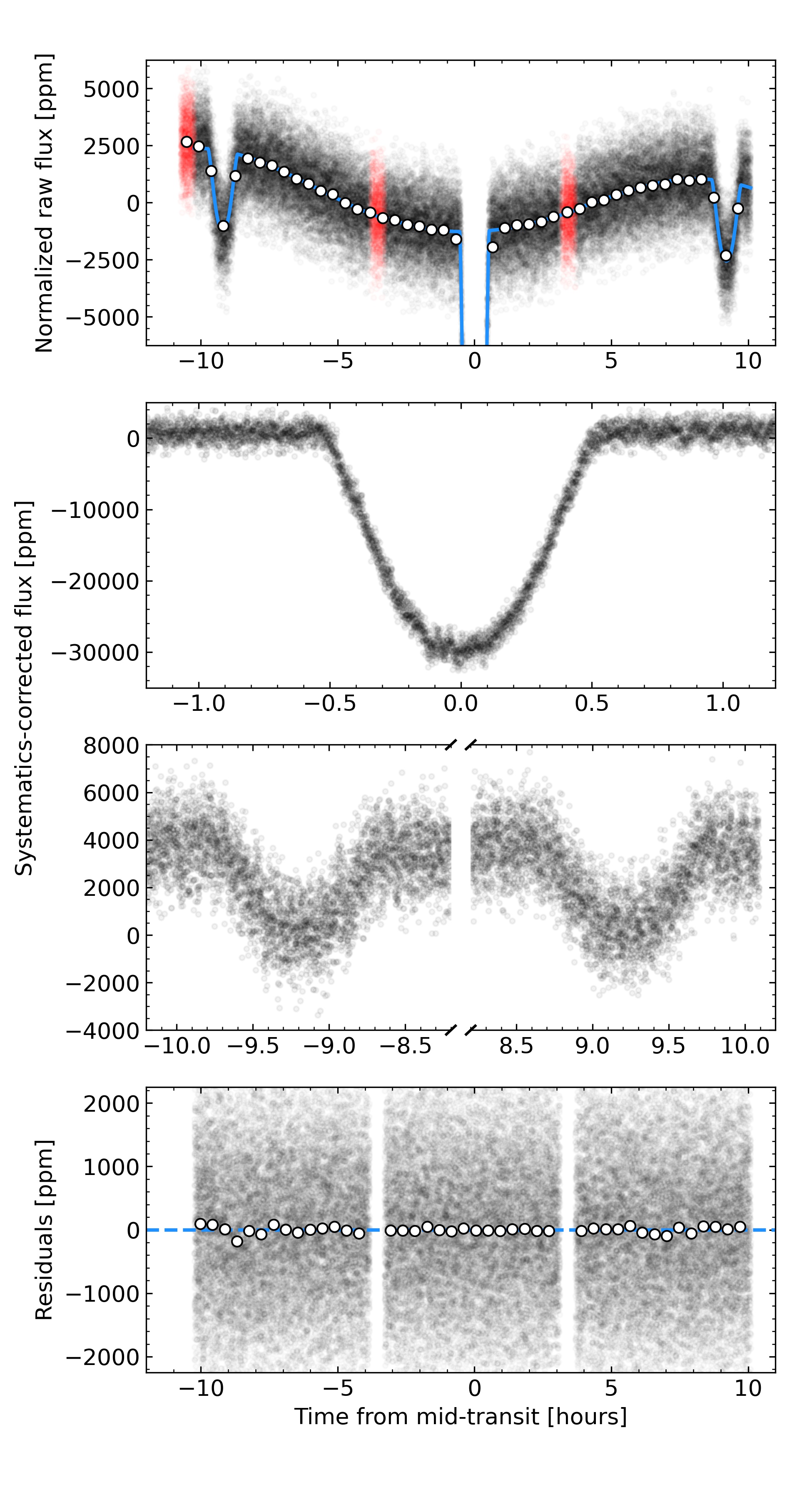}
    \vspace{-10mm}
    \caption{Broadband ($\lambda\geq2\,\mu$m) light-curve fit. The first panel shows the raw system flux as black dots, with the integrations not considered in the fit indicated by the red dots. We also show the median fit (blue line), along with the data binned at intervals of 27 minutes (1,000 integrations per bin; white circles). The second and third panels show the systematics-corrected data zoomed-in on the transit and secondary eclipses, respectively. The fourth panel shows the residuals from the best-fit to the data.}
    \label{fig:wlc_fit}
\end{figure}

\begin{figure*}
    \centering
    \includegraphics[width=1.\linewidth]{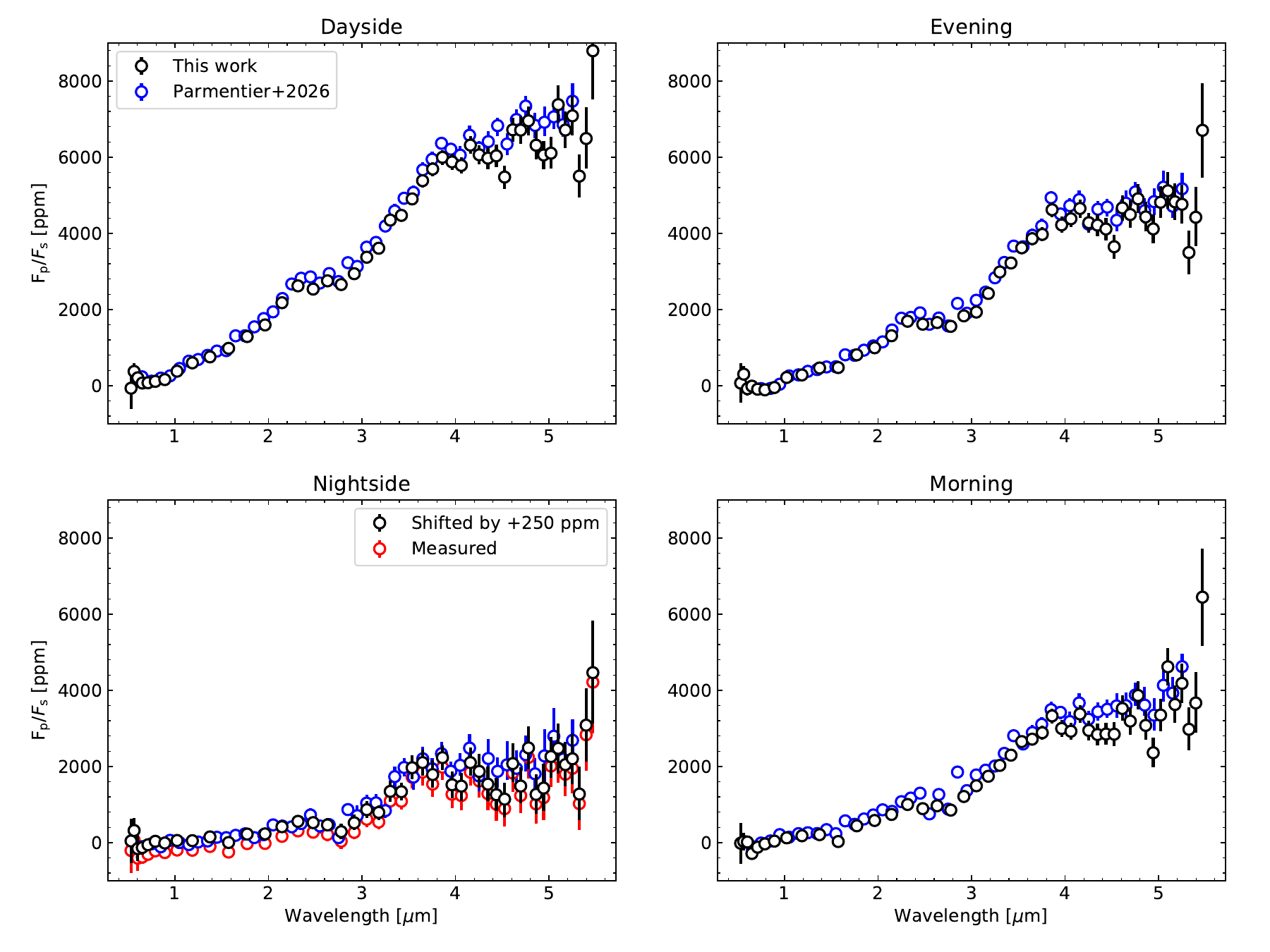}
    \caption{Planet-to-star flux ratio spectra of NGTS-10\,Ab (black circles) extracted at orbital phases $-180^\circ$ (nightside), $-90^\circ$ (evening), $0^\circ$ (dayside), and $90^\circ$ (morning) and shown on the same vertical scale. For the nightside spectrum, we show the measured data in red, along with the data that has been offset by $+250$\,ppm to avoid negative flux values. The spectra presented in \citet{Parmentier2026} are shown for comparison.}
    \label{fig:FpFs_spectra_4_phases}
\end{figure*}

\begin{figure}
    \centering
    \includegraphics[width=1.\linewidth]{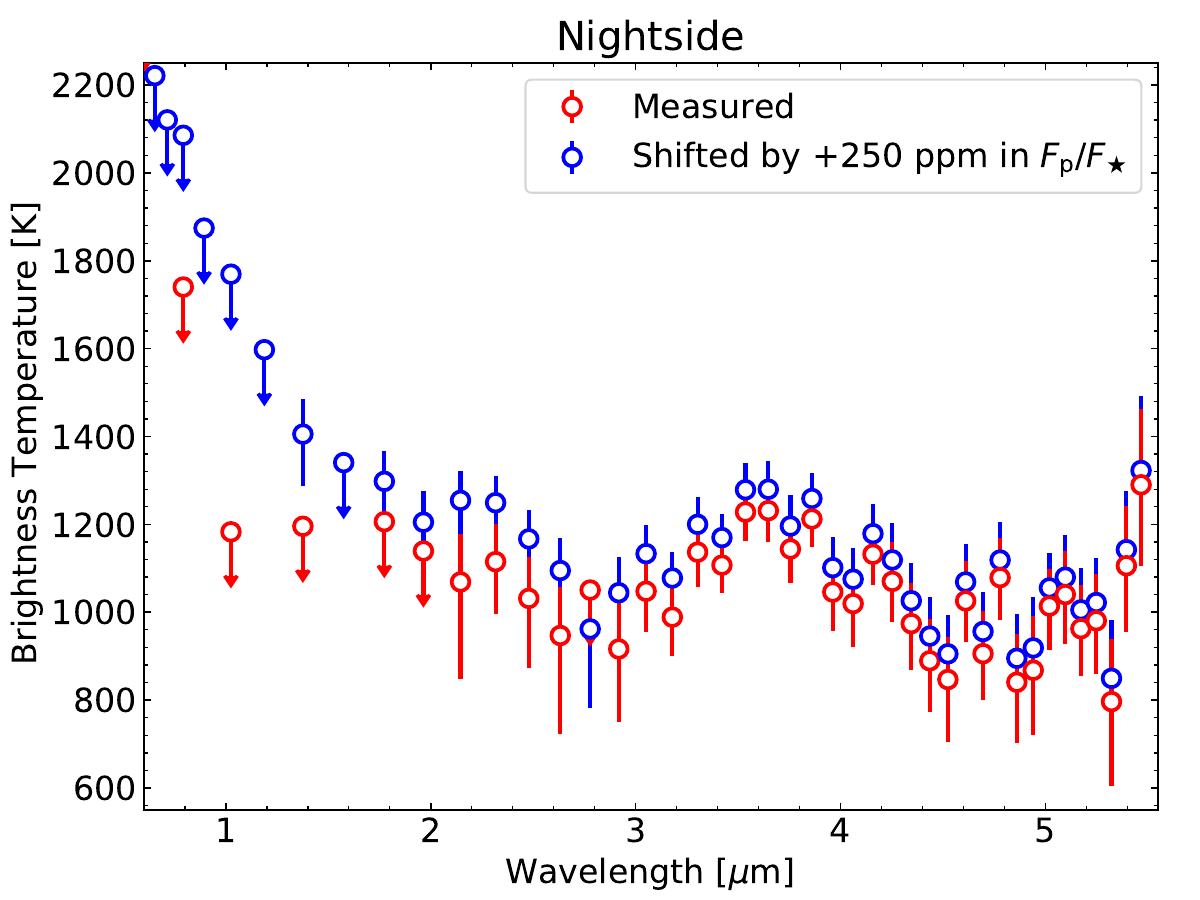}
    \caption{Impact of the applied DC offset (250\,ppm in $F_\mathrm{p}/F_\star$ space) to the nightside brightness temperature spectrum of NGTS-10\,Ab. We compare the spectrum obtained directly from the light curve fitting (red points) to the one shifted by +250\,ppm (blue points). Data points that are less than 1$\sigma$ above 0 in $F_\mathrm{p}/F_\star$ space are shown as $2\sigma$ upper limits. At short wavelengths, a few points from the non-shifted spectrum are missing as they are more than 2$\sigma$ below 0. The short wavelength data are primarily affected by the offset, whereas the relative change in $F_\mathrm{p}/F_\star$ at longer wavelengths ($>3\,\mu$m) results in a 40--90\,K difference in brightness temperature. We note that the steady rise in brightness temperature upper limits at short wavelengths should not be interpreted as an increase in planetary emission, but rather as a decrease in the brightness temperature sensitivity.}
    \label{fig:nightside_wo_offset}
\end{figure}

\begin{figure}
    \centering
    \includegraphics[width=1.\linewidth]{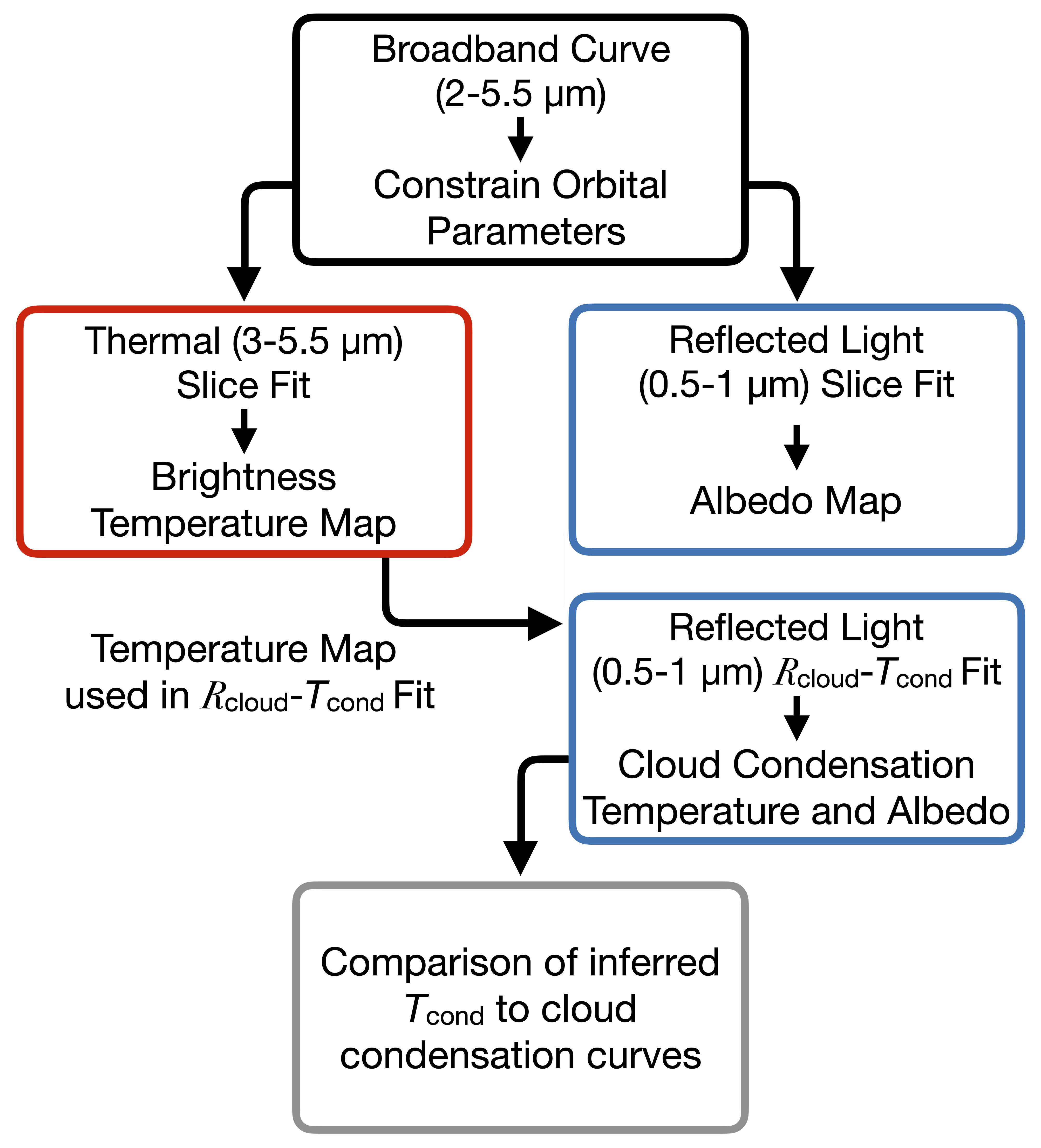}
    \caption{Flowchart of the thermal and reflected light phase curve fitting presented in Section \ref{sec:method}. We first fit the broadband ($\lambda = 2$--$5.5\,\mu$m) to constrain the system's orbital parameters that are then fixed at the thermal/reflected light phase curve fitting stage. After the broadband light curve fit, we perform the thermal and reflected light fits using the slice parametrization, where the thermal emission and reflectance longitudinal distributions of the planet are inferred without assuming a functional form for the maps. In a third step, we use the brightness temperature distribution from the thermal slice fit to model the reflectance distribution assuming that the planet is covered by clouds of uniform reflectance $\mathcal{R_\mathrm{cloud}}$ which evaporate above a temperature $T_\mathrm{cond}$. Finally, we infer the cloud species, as well as their condensation altitudes, that are consistent with the inferred condensation temperature and the gradient of the temperature-pressure profile.}
    \label{fig:flowchart}
\end{figure}

\begin{figure*}
    \centering
    \includegraphics[width=0.9\linewidth]{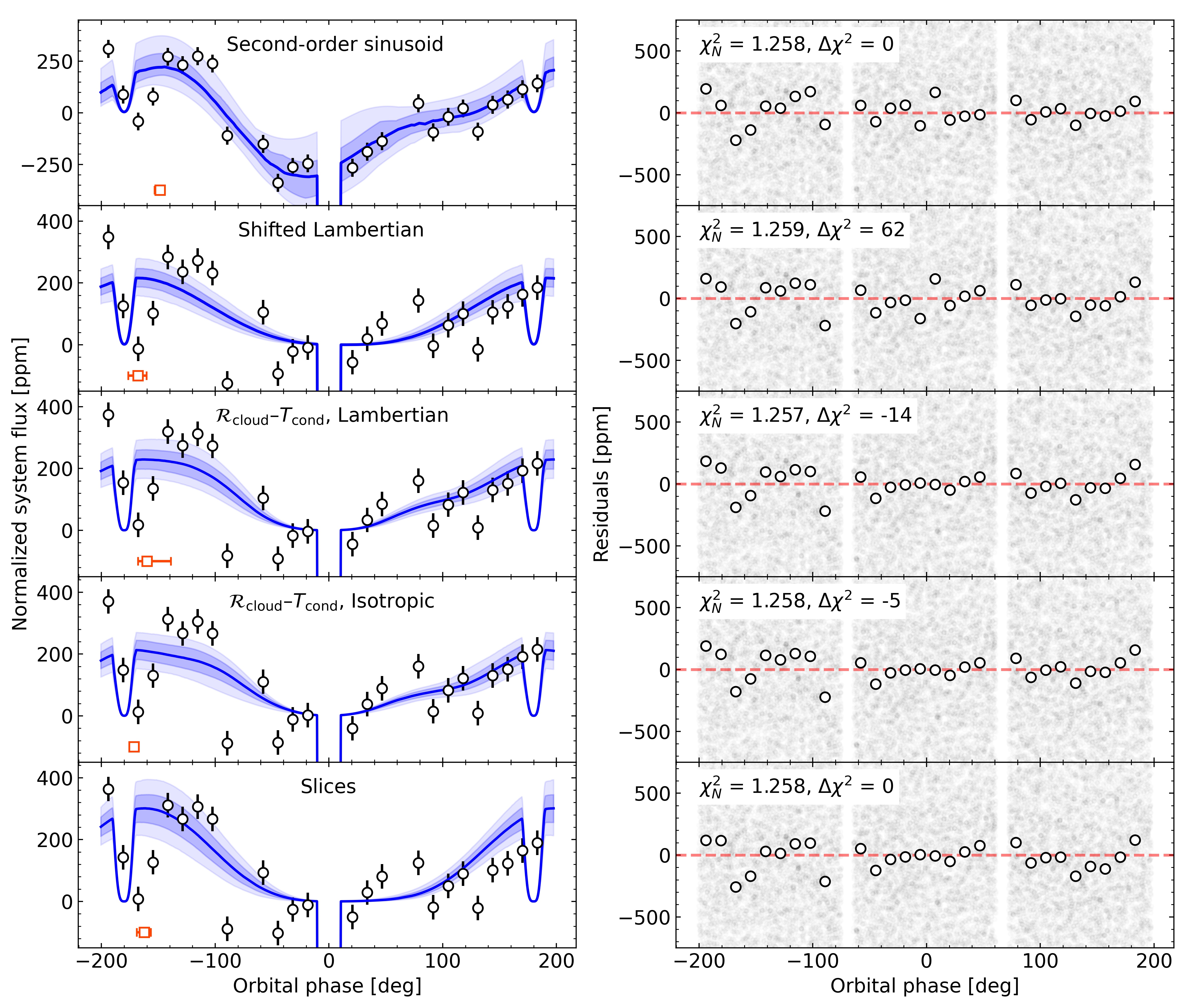}
    \caption{Comparison of the systematics-corrected reflected light phase curve fits for the five parameterizations tested. Left: The data (black circles) are binned at intervals of 40 minutes (1,500 integrations per bin) for visual clarity. The median fit (blue line) and the 1/2$\sigma$ confidence intervals (shaded regions) are shown. In contrast with the other parameterizations, the second-order sinusoid allows for negative phase curve flux, and reaches values of $-250$\,ppm near primary transit. The inferred phase curve offsets are indicated by the orange squares, and are all found to occur after secondary eclipse (westward offsets). Right: Residuals of the fits (black dots), binned at intervals of 40 minutes (black circles). The reduced chi-square and relative chi-square ($\Delta\chi^2$; relative to the second-order sinuoid model) values of each model fit are shown. The value assumed for the uncertainties in the chi-square is the point-to-point scatter. The $\mathcal{R}_\mathrm{cloud}$--$T_\mathrm{cond}$ Lambertian fit produces the lowest chi-square ($\Delta\chi^2$ = -14), but all models result in virtually the same reduced chi-square.}
    \label{fig:refl_light_fits}
\end{figure*}

\begin{figure*}
    \centering
    \includegraphics[width=1.\linewidth]{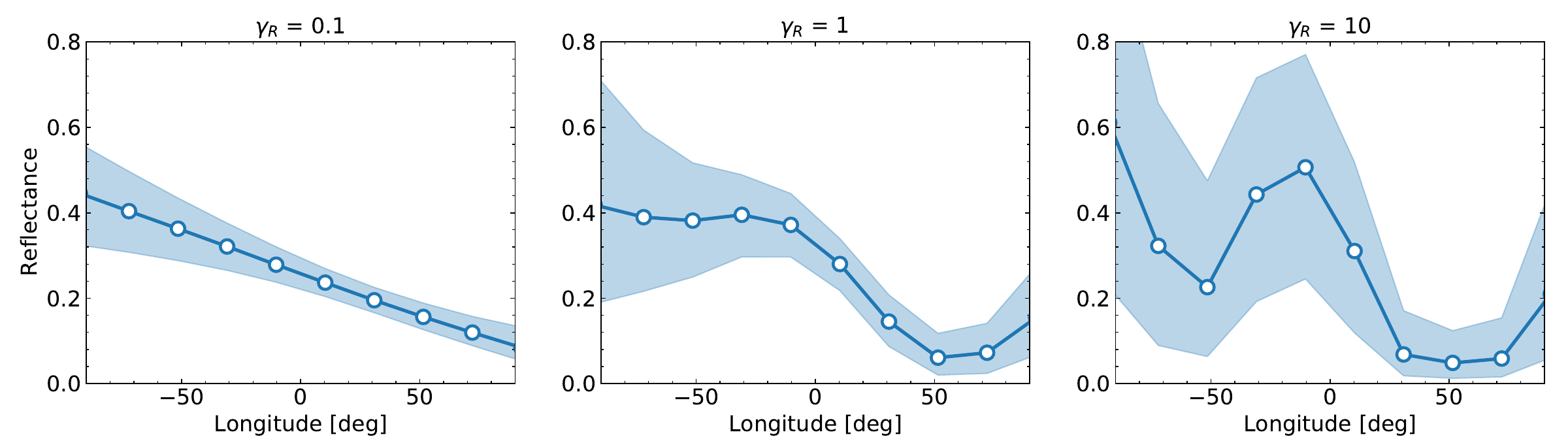}
    \caption{Impact of the regularization strength on the retrieved reflectance distribution from the reflected light phase-curve fit for values of $\gamma_R$ = 0.1 (left), 1 (middle), and 10 (right). Lower values of $\gamma_R$ correspond to stronger penalization against second derivatives in the reflectance map, resulting in a linear distribution when the penalization is strong ($\gamma_R=0.1$; left) and a solution that shows significant oscillations when the penalization is weak ($\gamma_R = 10$; right). The intermediate solution ($\gamma_R = 1$; middle) provides a compromise between both extremes.}
    \label{fig:RL_maps_vs_gamma}
\end{figure*}


\begin{figure}
    \centering
    \includegraphics[width=1.\linewidth]{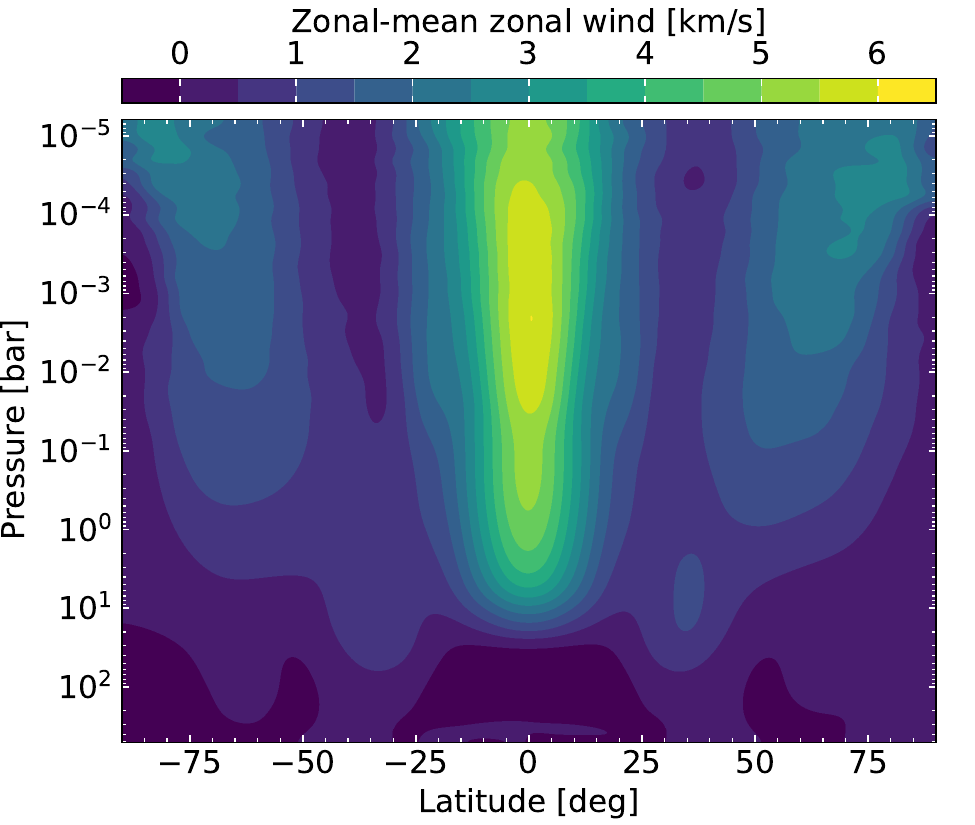}
    \caption{Longitudinal average of the zonal wind speed as a function of latitude and pressure for the $r_\mathrm{part} = 5\,\mu$m cloudy GCM.}
    \label{fig:zonal_mean_wind}
\end{figure}

\begin{figure*}
    \centering
    \includegraphics[width=0.9\linewidth]{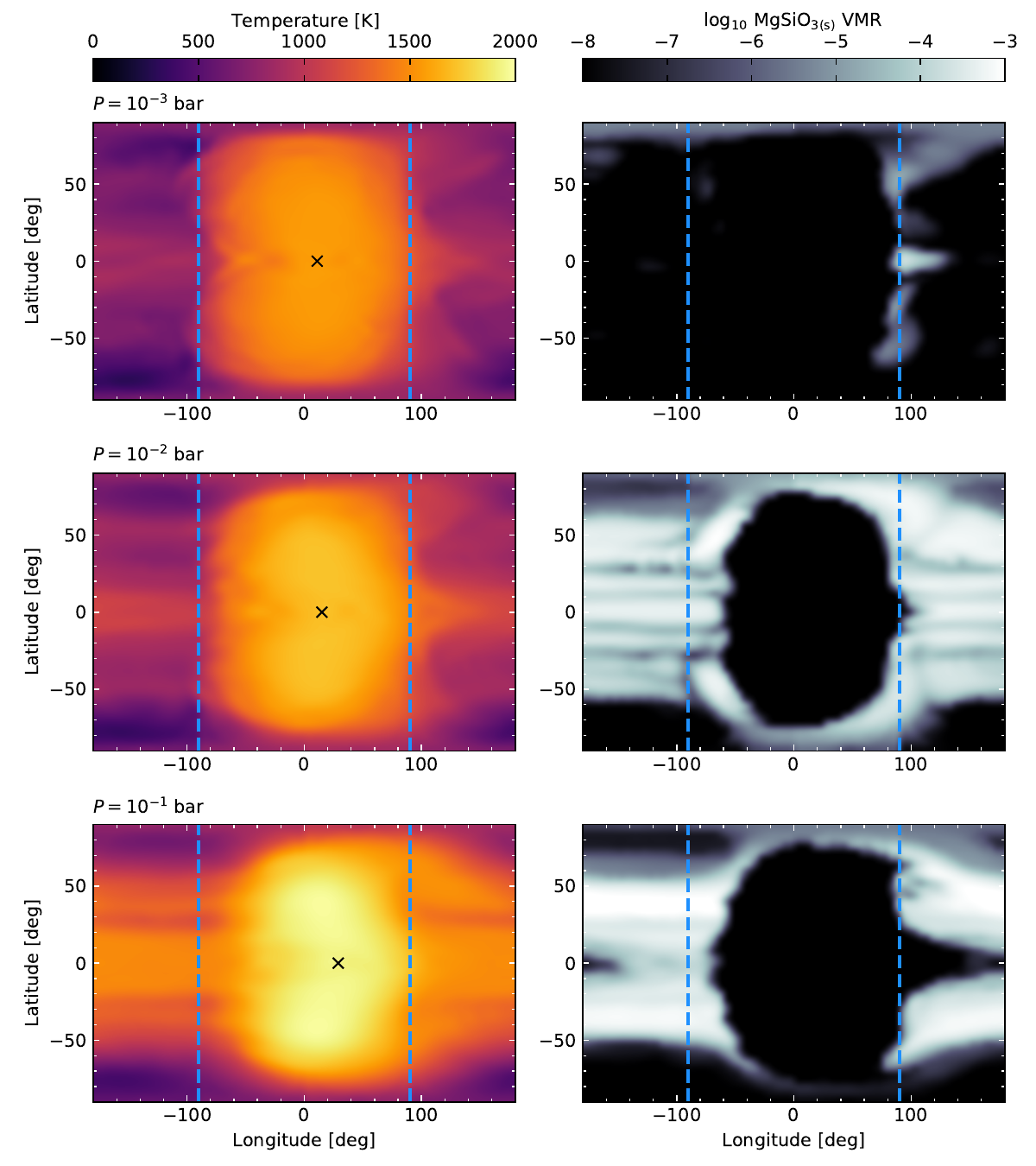}
    \caption{Temperature (left) and MgSiO$_3$ cloud tracer distributions from the $r_\mathrm{part} = 5\,\mu$m cloudy GCM, shown at pressures of $10^{-3}$\,bar (top), $10^{-2}$\,bar (middle), and $10^{-1}$\,bar (bottom). The terminator region ($\varphi=\pm90^\circ$) is indicated by the blue dashed lines. The crosses indicate the longitude at which the hemisphere-averaged emission ($T^4$) reaches its peak.}
    \label{fig:GCM_maps}
\end{figure*}

\begin{figure}
    \centering
    \includegraphics[width=1.\linewidth]{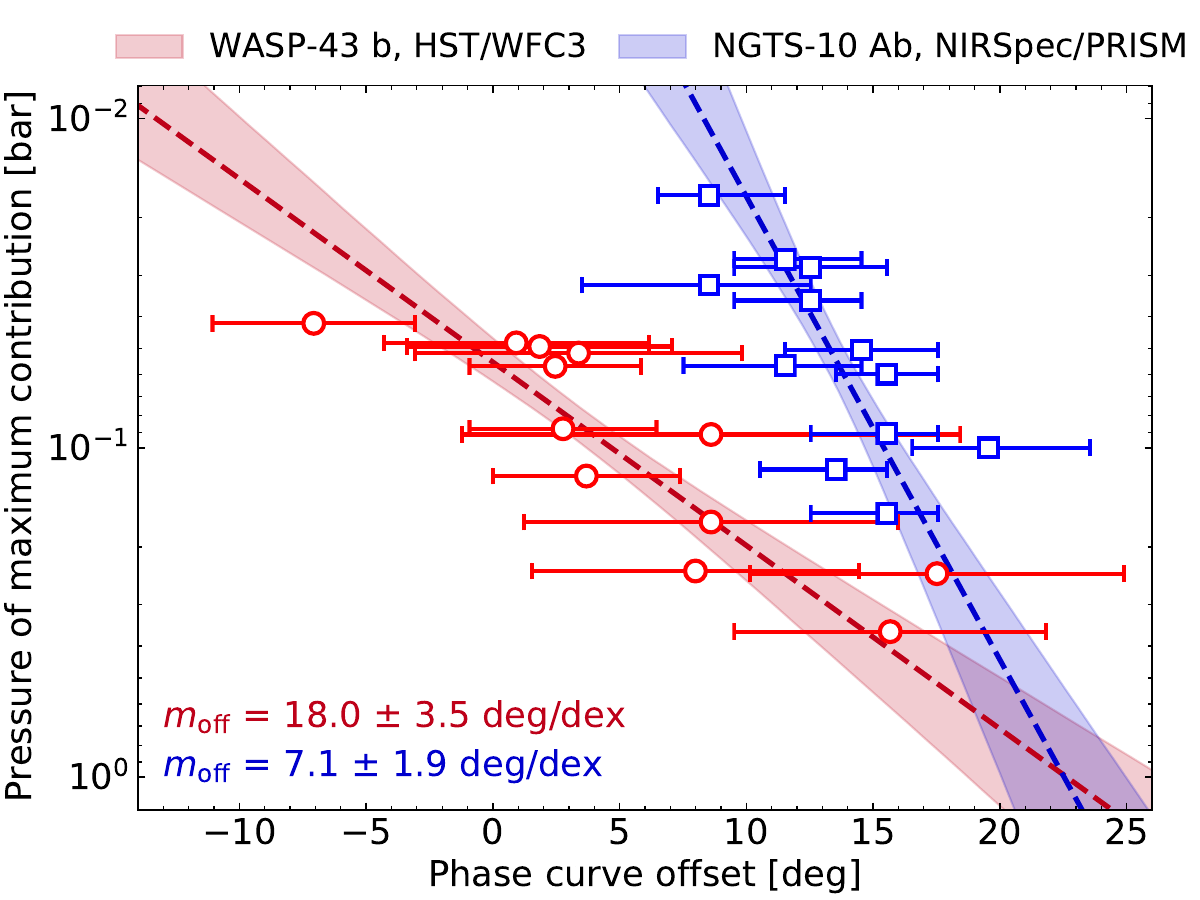}
    \caption{Offsets versus pressure values of WASP-43\,b as measured from its HST/WFC3 phase curve (red circles) and contribution function presented in \citet{Stevenson2014}. Following the same methodology as in Figure \ref{fig:offs_vs_press}, we measure a slope of $18.0\pm3.5$\,degrees per pressure dex (red dashed line and shaded region), $\sim$2.5 times steeper than the slope measured for NGTS-10\,Ab (blue dashed line and shaded region). The offset versus pressure values of NGTS-10\,Ab (blue squares) are shown for comparison.}
    \label{fig:offs_vs_pressure_W43b}
\end{figure}

\begin{figure}
    \centering
    \includegraphics[width=1.\linewidth]{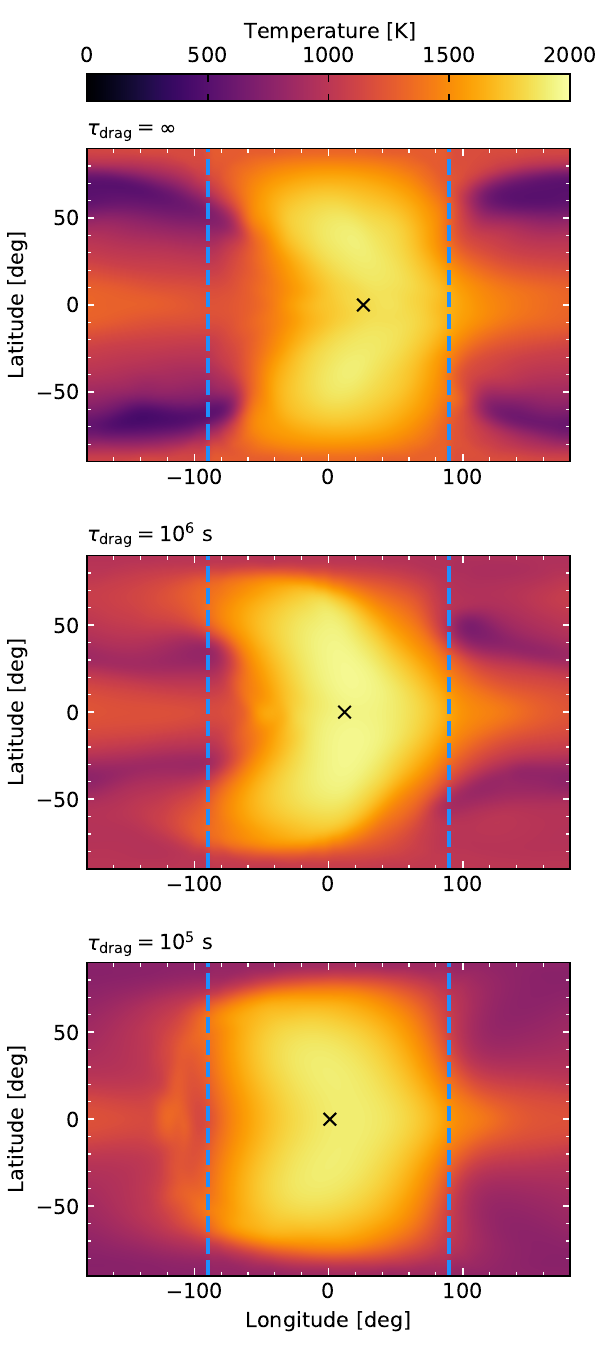}
    \caption{Temperature distributions from the cloudless GCMs for decreasing drag timescales of $\infty$ (the fiducial model), $10^6$\,s, and $10^5$\,s. The terminator region ($\varphi=\pm90^\circ$) is indicated by the blue dashed lines. The crosses indicate the longitude at which the hemisphere-averaged emission ($T^4$) reaches its peak.}
    \label{fig:GCM_maps_drag}
\end{figure}

\clearpage




\clearpage
\bibliography{main}



\end{document}